\documentclass[aps,pra,showpacs,notitlepage,floatfix,
twocolumn,superscriptaddress]{revtex4-1}
\usepackage{bbm}
\usepackage{mathrsfs}
\usepackage{subcaption}
\usepackage{epsfig}
\usepackage{graphicx}
\usepackage{amsfonts}
\usepackage[figuresright]{rotating}
\usepackage{amssymb}
\usepackage{amsmath}
\usepackage{dcolumn}
\usepackage{bm}
\usepackage{braket}
\usepackage[normalem]{ulem}

\usepackage[colorlinks,linkcolor=blue,anchorcolor=blue,citecolor=blue,urlcolor=blue]{hyperref}

\def\be{\begin{equation}} \def\ee{\end{equation}}
\def\bea{\begin{equation}\begin{aligned}} \def\eea{\end{aligned}\end{equation}}

\begin{document}

\title{Product Spectrum Ansatz and the Simplicity of Thermal States}

\author{John Martyn}
\affiliation{Department of Physics, University of Maryland, College Park, MD 20742, USA}
\author{Brian Swingle}
\affiliation{Condensed Matter Theory Center, Maryland Center for Fundamental Physics, Joint Center for Quantum Information and Computer Science, and Department of Physics, University of Maryland, College Park, MD 20742, USA}\affiliation{Institute for Advanced Study, Princeton, NJ, 08540}

\begin{abstract}
Calculating the physical properties of quantum thermal states is a difficult problem for classical computers, rendering it intractable for most quantum many-body systems. A quantum computer, by contrast, would make many of these calculations feasible in principle, but it is still non-trivial to prepare a given thermal state or sample from it. It is also not known how to prepare special simple purifications of thermal states known as thermofield doubles, which play an important role in quantum many-body physics and quantum gravity. To address this problem, we propose a variational scheme to prepare approximate thermal states on a quantum computer by applying a series of two-qubit gates to a product mixed state. We apply our method to a non-integrable region of the mixed field Ising chain and the Sachdev-Ye-Kitaev model. We also demonstrate how our method can be easily extended to large systems governed by local Hamiltonians and the preparation of thermofield double states. By comparing our results with exact solutions, we find that our construction enables the efficient preparation of approximate thermal states on quantum devices. Our results can be interpreted as implying that the details of the many-body energy spectrum are not needed to capture simple thermal observables.
\end{abstract}

\maketitle

\section{Introduction}
The preparation of non-trivial quantum states is one of the central challenges of quantum simulation and computation. Even given the kind of control envisioned in a full scale fault tolerant quantum computer, it is often not known what sequence of operations will realize a particular physical state of interest. A prime example is a ground state of a local Hamiltonian, or more generally, a thermal mixed state of a local Hamiltonian. On complexity theoretic grounds, there is likely no general purpose algorithm to efficiently prepare such states (for a review, see~\cite{nielsen2010quantum,riera2012thermalization,kempe2004complexity}), but we expect it to be possible in many cases of physical interest.

For example, if the system and temperature are such that the thermal state can be cast as an approximate quantum Markov state with finite correlation length, then there exists an efficient preparation procedure (although determining the procedure may still be hard)~\cite{swingle2016mixed,kato2016quantum,brandao2016finite}. More broadly, many approaches have been proposed, some applicable to special models and others general but possibly requiring a very long computation time~\cite{Temme2009quantummetro,Bilgin2010quantumgibbs1d,Chowdhury2016quantumgibbs,Kastoryano2014commutinggibbs,Brandao2017sdp}.

Perhaps the most natural algorithm to prepare a thermal state is to weakly couple the system of interest to a heat bath of the appropriate temperature and wait for equilibrium to be reached. As amply demonstrated in nature, this algorithm is often successful, but it is not fully satisfactory for our purposes. Even assuming we have access to the required heat bath (which depends on the physical setup), the approach to equilibrium may be slow (and it may be hard to tell if equilibrium has been reached) if the bath is inefficient at thermalizing the system, see e.g. Reference~\cite{terhal2000problem}.

The bath approach is particularly problematic if we want to prepare a special purification of the thermal state known as a thermofield double state~\cite{Takahashi:1996zn}. This state is defined on two copies of the system, where the second copy functions as a kind of minimal heat bath which can thermalize the original system (typically, the heat bath is much larger than the system). Whereas the typical state of the system-bath composite is highly complex after thermal equilibrium is reached, the thermofield double has a particularly orderly and simple kind of system-bath entanglement. Distilling this orderly pattern of entanglement from a general complex system-bath state is a non-trivial task.

Thermofield double states are central to many modern developments in quantum many-body physics, so it is desirable to be able to prepare them. For example, these states play an important role in quantum gravity in anti de Sitter space where they translate, via the AdS/CFT correspondence, to special geometries consisting of two entangled black holes \cite{Israel:1976ur,maldacena2003eternal}. Recent discussions of black hole firewalls, teleportation through wormholes, and other exotica can be naturally setup using the thermofield double state \cite{maldacena2018eternal, Almheiri:2012rt, Almheiri:2013hfa,susskind2016er, Gao:2016bin, maldacena2017diving}. To investigate these phenomena and their analogs in quantum simulations of quantum gravity and other many-body systems, it would be useful to be able to prepare thermofield double states.

In this work, we propose a method to prepare approximations of thermal mixed states and thermofield double states on a quantum device. The basic method we propose assumes the ability to apply arbitrary two-qubit unitaries, naively requiring something like the versatility of a quantum computer, but refinements to more limited devices should be possible. The approach is variational: given a suitable class of mixed states, we minimize the free energy of the system over the class. The free energy consists of two parts, an energy term and an entropy term. The energy is either physically measured or, in a version of our approach adapted to classical simulation, calculated on a computer. The entropy is calculated from the state, with the class of mixed states we consider being specially chosen to make this possible. Given a procedure to prepare an approximation of a thermal mixed state, our approach also immediately gives a procedure to prepare an approximation to the corresponding thermofield double state.

\begin{figure}
  \centering
  \includegraphics[width=.48\textwidth]{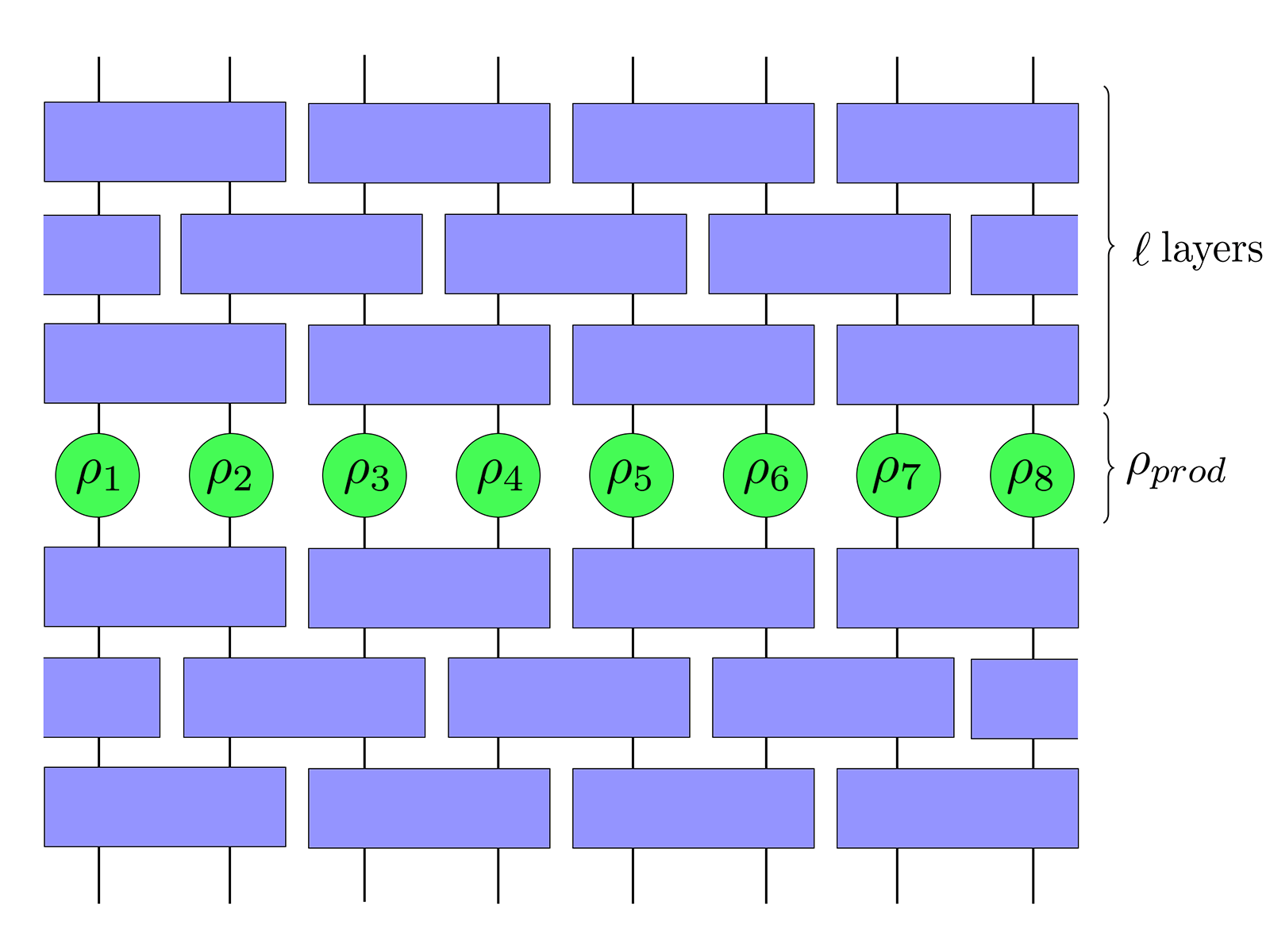}
  \caption{A tensor network depiction of the PSA on a system of 8 qubits. $\rho_{\mathrm{prod}}$ is the product spectrum state to which $\ell$ layers of 2-qubit unitaries are applied. In this diagram, $\ell=3$.} \label{fig:psa}
\end{figure}

In somewhat more detail, our work consists of two components. First, we propose that many thermal states of interest can be approximated by a class of mixed states which we call product spectrum states. Second, we show that the free energy of product spectrum states can be efficiently computed and construct a variational procedure to obtain said approximation. For spin systems, product spectrum states are defined in Figure~\ref{fig:psa}. A set of spins are prepared in an uncorrelated thermal state, $\rho_{\mathrm{prod}}$, in which each spin has a probability to be up or down independent of all others. Then a sequence of unitary transformations is applied that entangles the spins together without changing the eigenvalue spectrum. It follows that the entropy of the system can be computed from the individual probabilities of the initial uncorrelated spins. Moreover, because it is straightforward to purify an uncorrelated thermal state, we also immediately obtain a recipe for preparing a corresponding thermofield double state.

Since the eigenvalue spectrum of the many-body state is the product of the spectra of each independent spin, we call our construction the product spectrum ansatz (PSA). Aspects of our approach are similar to recent work reported in References \cite{pachos2018quantifying, kato2016quantum}; two crucial differences are that they did not discuss the thermofield double state and that they assume full knowledge of the many-body spectrum. We also note that, as elaborated in the concluding discussion, the product spectrum ansatz is just the first approximation in a systematically improvable hierarchy which we call the Markov spectrum ansatz.

The physical idea behind the PSA, at least for quantum chaotic systems, is the following. The full many-body spectrum of a thermal state of a quantum chaotic system is highly complex, even random-matrix-like~\cite{PhysRevLett.52.1}, as can be demonstrated explicitly for small system sizes, e.g.~\cite{Cotler2016bhrmt}. However, this complexity is not expected to be visible in few-body physical observables, so perhaps the spectrum could be approximated by something much simpler while preserving the physics of interest. Similarly, although the unitary circuit which produces energy eigenstates from product states is highly complex, the actual mixed thermal state is often not, having in many cases short-range correlations and short-range entanglement. Hence, it might be possible to replace the highly complex energy eigenstates with much simpler states obtained from a low depth quantum circuit~\cite{swingle2016mixed}. There are also motivations for the PSA approach from the study of tensor networks and complexity~\cite{Swingle:2009bg,Susskind:2014rva,Stanford:2014jda,Brown:2015lvg,Chapman:2018hou} in holographic duality~\cite{Maldacena:1997re}.

Our precise results are as follows. We formally define the product spectrum ansatz and apply it to both a local spin model, the mixed field Ising chain, and a non-local fermion model, the Sachdev-Ye-Kitaev (SYK) model~\cite{Sachdev1993,kitaev2015,Polchinski:2016xgd,syk} (see References \cite{rosenhaus2018introduction,sarosi2017ads2} for reviews). For the SYK model, our procedure can be interpreted as a method to approximately prepare a short wormhole. In the local case, we compare the product spectrum ansatz to exact results in small size systems and show how the product spectrum ansatz can be extended to larger size systems by virtue of its local structure of correlations. In the non-local case, we again compare approximate and exact data for small size systems. We also compare our approach to an alternative method discussed in Reference~\cite{maldacena2018eternal} for preparing approximate thermofield double states of the SYK model. In both cases we find good agreement across a variety of couplings and temperatures.

As we discuss at length in the conclusions, there is ample room to improve on our scheme. These include refinements to the structure of the ansatz and improvements to the optimization method. For example, in the non-local case, the variational calculations are difficult to extend to larger size, primarily because the minimization of the free energy is slow. It would also be interesting to physically implement our method on a quantum device. We comment on the important question of noise resilience in the conclusions.

\section{Product spectrum ansatz}
\subsection{Formalism}
This section begins with a formal definition of the product spectrum ansatz. We focus on a system of $n$ spin-1/2 degrees of freedom for simplicity, but the extension to other kinds of systems is straightforward. A product spectrum state is defined by a set of single spin effective energies $\{\epsilon_r\}$, $r\in\{1,\cdots,n\}$ and an $n$-spin unitary $U$. The PSA density matrix is then
\begin{equation} \label{eq:psa_def}
  \rho_{\text{PSA}} = U \rho_{\mathrm{prod}} U^\dag = U \left[\bigotimes_{r=1}^n \frac{e^{-\beta \epsilon_r P_r}}{1+ e^{-\beta \epsilon_r}} \right] U^\dagger,
\end{equation}
where $T=1/\beta$ ($k_B = 1$) is the temperature and $P_r = \frac{I - \sigma^z_r}{2}$ is a projection onto the spin down state for spin $r$. The quantity inside the brackets is what we call the product spectrum because it takes the form of a thermal state for the decoupled effective spin Hamiltonian $\sum_r \epsilon_r P_r$. Since conjugating with the unitary $U$ does not change the eigenvalue spectrum, it follows that $\rho_{\text{PSA}}$ has the same spectrum as the product of single spin states, and that it is positive and normalized to unity.

Although the definition in Eq.~\eqref{eq:psa_def} in principle allows for any unitary $U$, in many cases we will further restrict $U$ to consist of a low-depth quantum circuit. For example, if the $n$ spins are arranged in a one-dimensional array, then we will take the unitary $U$ to consist of alternating layers of two-spin gates acting on even or odd pairs as in Figure~\ref{fig:psa}. Clearly the expressive power of the PSA approximation will increase as the depth of $U$ is increased, but it may also be harder to prepare and study high depth PSA states.

One important question is how the depth of $U$ scales with the temperature. If the ground state has some entanglement, then there will be some net increase in depth going from infinite temperature to zero temperature. However, it is far from clear what the precise dependence is, or even if the depth depends monotonically on temperature. If a state has a correlation length $\xi$, then the depth should be lower bounded by a constant times $\xi$. If the Hamiltonian is gapped, then the correlation length does not diverge at zero temperature and we may expect the depth to be of order the correlation length. However, this is really only a lower bound, since topological phases require a depth proportional to system size at zero temperature despite having short-range local correlations. If Hamiltonian is gapless and the correlation length diverges at low temperature, then the needed depth will diverge as well.

What follows is a theoretical justification for the PSA for a class of chaotic, thermalizing quantum systems which are assumed to obey the eigenstate thermalization hypothesis (ETH)~\cite{Deutsch1991ETH,Srednicki1994ETH}. Since these systems can in principle be interacting and arbitrarily large, our construction justifies the PSA for interacting systems  and systems in the thermodynamic limit. The argument proceeds in stages. Suppose first that an arbitrary unitary $U$ is allowed in Eq.~\eqref{eq:psa_def}. Then $U$ can be taken to diagonalize the Hamiltonian, and by sending the constants $\epsilon_r \rightarrow \pm \infty$, one can single out any particular energy eigenstate. Using ETH, this pure PSA state already reproduces all the local data of the thermal state at the corresponding temperature.

More generally, it is possible to choose the spectrum $\epsilon_r$ such that the weight of $\rho_{\text{PSA}}$ is concentrated on a set of exact energy eigenstates of the appropriate energy density for a temperature $T$. Every term in the mixture then reproduces the correct local energy density and the fact that the state is a mixture generates entropy, hence lowering the free energy, $F=E-TS$. The total number of states with energy density corresponding to temperature $T$ is of order $e^{S(T)}$, so the minimum free energy in this kind of state is indeed $F_{\text{PSA}} \approx E(T)- T S(T)$, the thermal value.

The remaining question is then what happens if the complexity of $U$ is restricted. Clearly, we can no longer use energy eigenstates in the construction. However, the use of such states is intuitively not necessary, since the actual thermal state typically has short-range entanglement (instead of extreme long-range entanglement masquerading as thermal physics as in an energy eigenstate). It can be shown using quantum information methods that many thermal states ranging from non-interacting particle states to strongly interacting plasma states holographically dual to black holes have a local structure such that the global state can be reconstructed from local data \cite{swingle2016mixed}, and this locality property implies that the thermal state can be obtained as a mixture of low-complexity states. Hence, we argue that for many thermal states of interest, the complexity of $U$ can also be taken to be relatively low, much less than the complexity of the Hamiltonian-diagonalizing unitary.

Of course, it should be mentioned that the ground state and the infinite temperature state have exact representations as PSA states. If the ground state is pure, it can be approximated by a circuit of some depth acting on a product state, and can therefore be described by the PSA. Likewise, since the infinite temperature state is maximally mixed, it is a product of single qubit maximally mixed states, which can also be described by the PSA. Moreover, it is also conjectured on physical grounds and even known rigorously in some cases, that the corresponding unitary $U$ need not be deeper than linear in system size for a broad class of physical systems. We note in passing that many kinds of integrable systems should also be well approximated by PSA states.

\subsection{Benchmarking Results}
The validity of the PSA can be studied numerically by determining the optimal parameters $\{ \epsilon_r \}$ and 2-qubit gates in $U$ (fixing the number layers) such that the PSA state best approximates the thermal state of interest. We implement the PSA by varying over the spin effective energies $\{\epsilon_r\}$ and the 2-qubit gates to search for a minimum of the free energy. In a classical simulation, this variation is done layer by layer. It is most efficient to vary over the 2-qubit gates one at a time, and continually repeat this process until a satisfactory minimum is attained. This procedure may be performed by a gradient descent method, or by an iterative algorithm such as that described in Reference \cite{PhysRevB.79.144108}. There is no guarantee that the output of this procedure is the true global minimum of the PSA free energy.

\subsubsection{Mixed Field Ising Model}
We aim to establish the accuracy of the PSA on a local spin chain model. The candidate that we will examine is the mixed field Ising model with $n$ spins:
\begin{equation}
H_{\mathrm{Ising}} = -J \sum_{i=1}^{n-1} \sigma_i^z \sigma_{i+1}^z - h_z \sum_{i=1}^{n} \sigma_i^z - h_x \sum_{i=1}^{n} \sigma_i^x.
\end{equation}
In our notation, $\sigma^x_i$ and $\sigma^z_i$ are the Pauli operators at site $i$, and $h_x$ and $h_z$ are external magnetic field strengths. We will set $J=1$ and study the PSA at different values of the couplings $h_x$ and $h_z$. Since the PSA is intrinsically local, we believe that it should replicate the behavior of this model quite well with a few layers of unitaries applied to the product state.

\begin{figure}[h]
\begin{subfigure}{.24\textwidth}
  \centering
  \includegraphics[width=4.6cm]{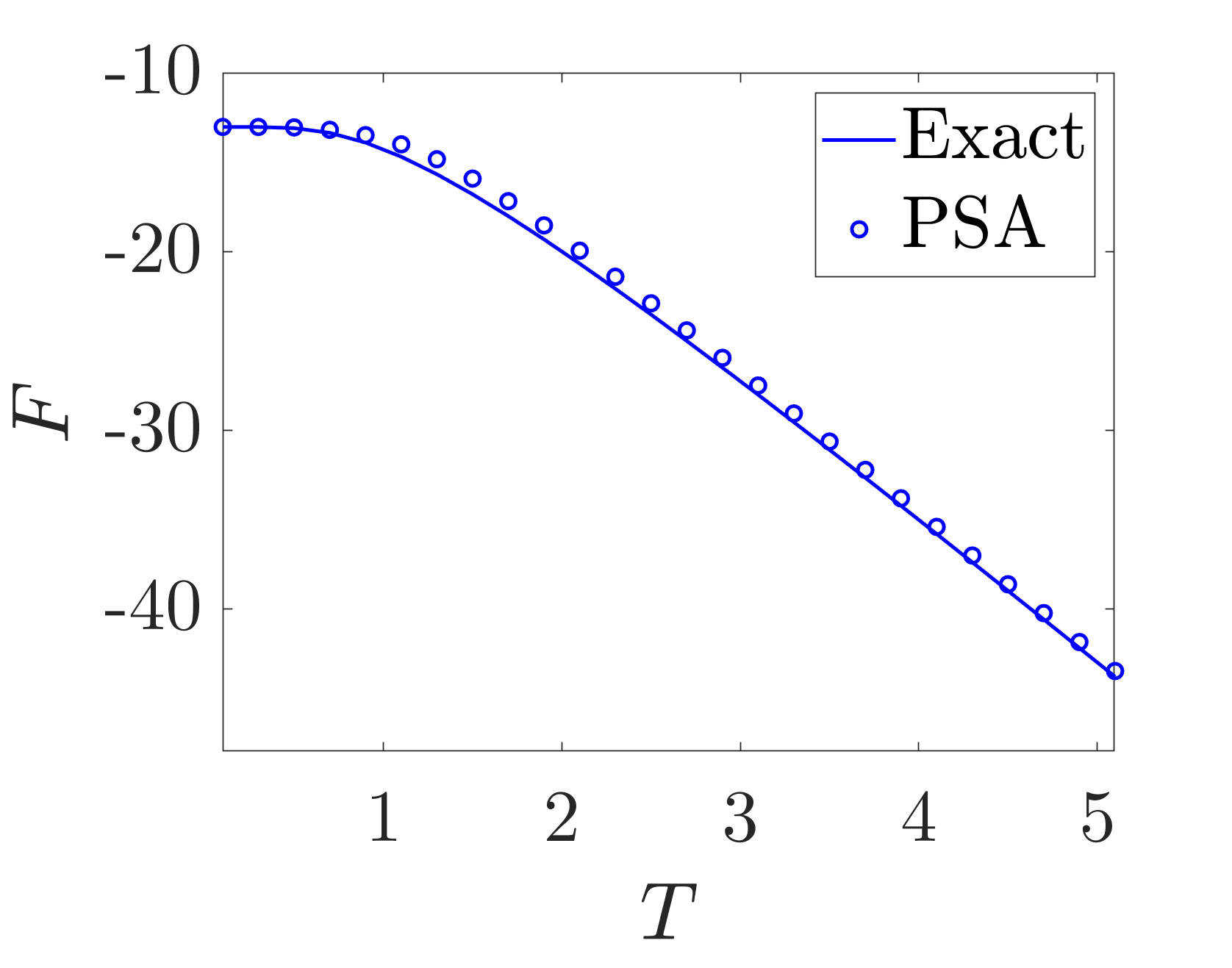}
  \caption{$h_x = 0.5,\ h_z = 0.1$}
  \label{fig:FreeEnergy(0.5,0.1)}
\end{subfigure}
\begin{subfigure}{.24\textwidth}
  \centering
  \includegraphics[width=4.6cm]{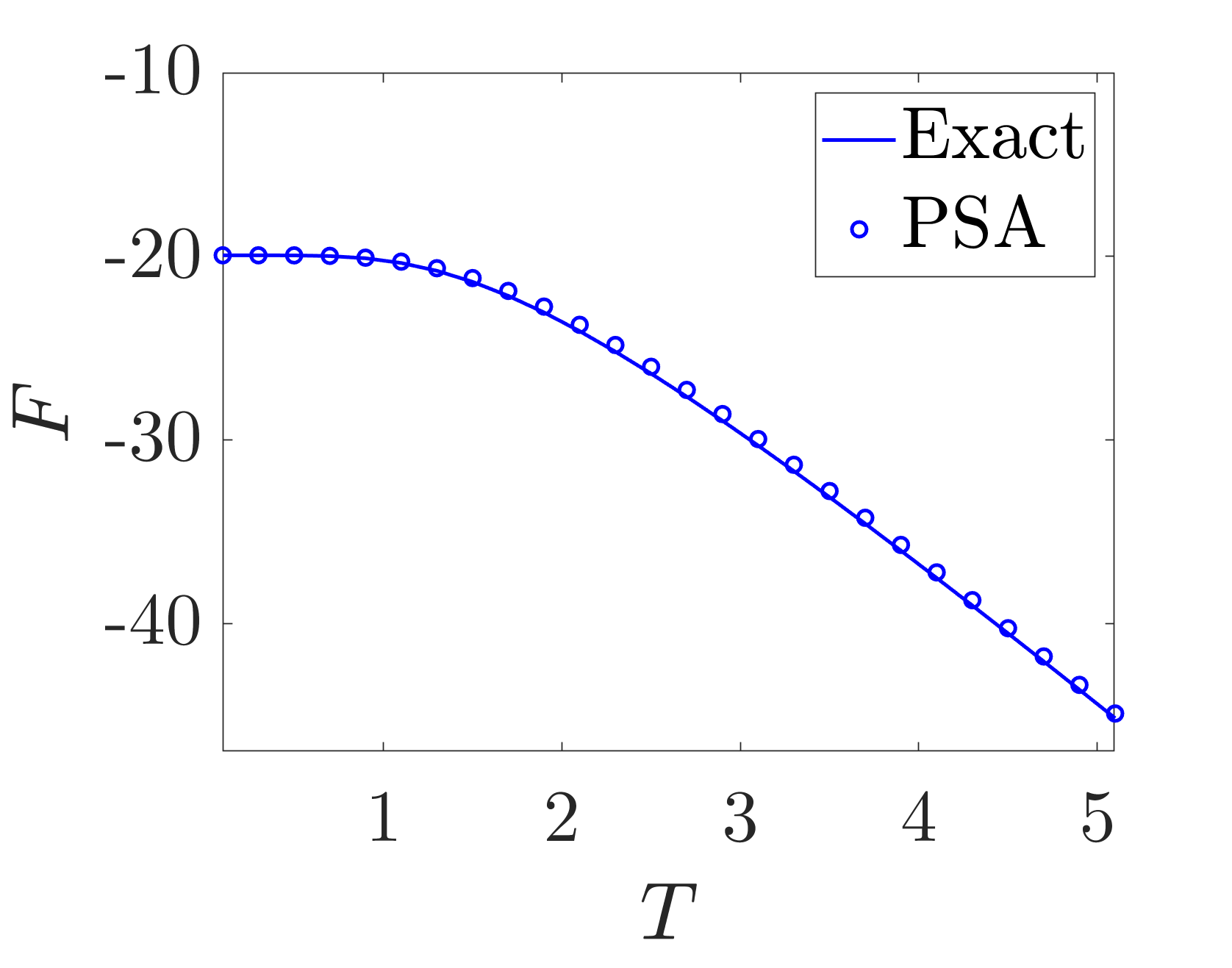}
  \caption{$h_x = 1.05,\ h_z = 0.5$}
  \label{fig:FreeEnergy(1.05,0.5)}
\end{subfigure}
\caption{Plots of exact and PSA free energy over the temperature range $T \in [0.1, 5.1]$ with $n=12$ spins and $\ell = 4$ layers of unitaries applied to the product spectrum.}
\label{fig:FreeEnergy}
\end{figure}

A comparison of the exact free energy and that attained with the PSA provides a means for probing the validity of the PSA in this scenario. We expect these free energies to take similar values over a range of temperatures and couplings, indicative of the accuracy of the PSA. Since determining the exact solution to this model for arbitrary $h_x$ and $h_z$ requires exponential resources, we consider a small chain with $n=12$ spins, such that we can compare the exact free energy to the PSA free energy. We focus on the scenarios $(h_x = 1.05,\ h_z=0.5)$ and $(h_x = 0.5,\ h_z=0.1)$, at which points the mixed field Ising model is nonintegrable. Figure \ref{fig:FreeEnergy} displays the exact and PSA free energies over the range of temperatures $T\in [0.1, 5.1]$ with $\ell = 4$ layers of 2-qubit gates applied to the product spectrum.

As evidenced by the plots in Figure \ref{fig:FreeEnergy}, the PSA free energy matches the exact free energy quite well over this range of temperatures.  The best agreement occurs in the regions of low and high temperature since the exact thermal state is a product spectrum state in the limits $T \rightarrow 0$ and and $T \rightarrow \infty$. The maximum discrepancy between the exact and PSA free energies in the $(h_x = 0.5,\ h_z = 0.1)$ case is $5.4\%$ at $T=1.3$, and that in the $(h_x = 0.5,\ h_z = 0.1)$ case is $1.4\%$ at $T=2.1$. Away from these temperatures, the discrepancies between the free energies decrease significantly. These results demonstrate that the PSA provides an excellent variational scheme for calculating the free energy of a thermal state governed by local interactions.

\subsubsection{SYK Model}

Our next aim is to examine the behavior of the PSA on a nonlocal fermion model. To analyze such a scenario, we selected the SYK model with $N$ Majorana fermions:
\begin{equation}
H_{\text{SYK}} = \sum_{i_1>i_2>i_3>i_4=1}^N J_{i_1i_2i_3i_4} \chi_{i_1}\chi_{i_2}\chi_{i_3} \chi_{i_4}
\end{equation}
\begin{equation}
 \quad \{\chi_{i},\chi_{j} \} = \delta_{ij},\ \langle J_{i_1i_2i_3i_4} \rangle = 0,\ \langle J^2_{i_1i_2i_3i_4} \rangle = \frac{3! J^2}{N^3}.
\end{equation}
In our notation, $\chi_i$ is a Majorana fermion at site $i$. The coupling $J_{i_1 i_2 i_3 i_4}$ is a Gaussian distributed random variable with mean $0$ and variance $ 3! J^2/N^3$. We set $J=1$ in our analysis. To represent the model numerically, we use a standard Jordan-Wigner representation to write the $N$ fermions in terms of $N/2$ spins via $\chi_{2i-1} = 2^{-1/2} \sigma^x_i \prod_{j < i} \sigma^z_j$ and $\chi_{2i} = 2^{-1/2} \sigma^y_i \prod_{j < i} \sigma^z_j$. Our results are for single disorder realizations of the SYK model.

\begin{figure}[h]
  \centering
  \includegraphics[width=.48\textwidth]{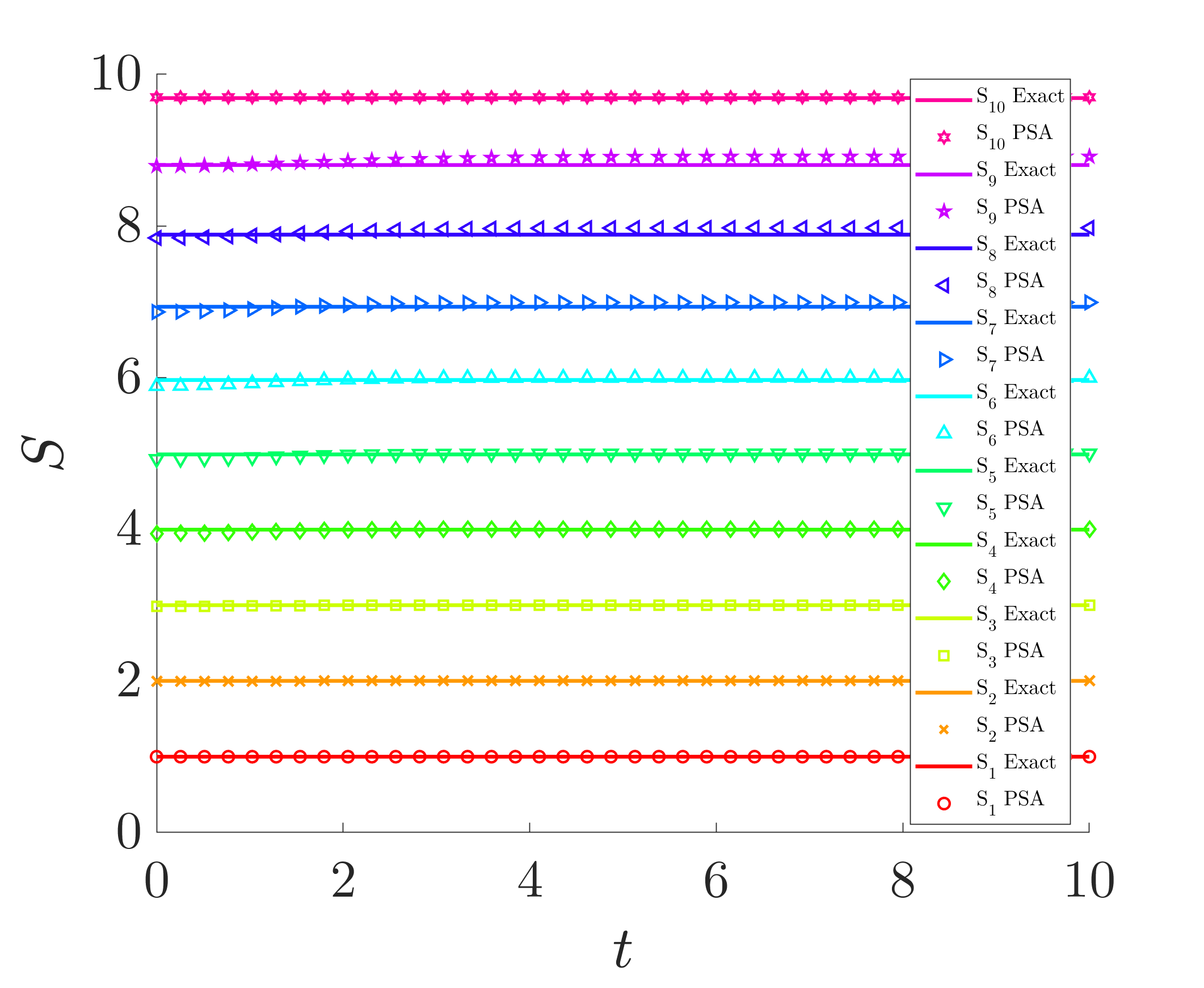}
  \caption{Exact and PSA subsystem entanglement entropies for the SYK Model with $N=20$, $T=0.7$, and $\ell=16$.} \label{fig:SYKEntropy}
\end{figure}

\begin{figure}[h]
\begin{subfigure}{.245\textwidth}
  \centering
  \includegraphics[width=4.3cm]{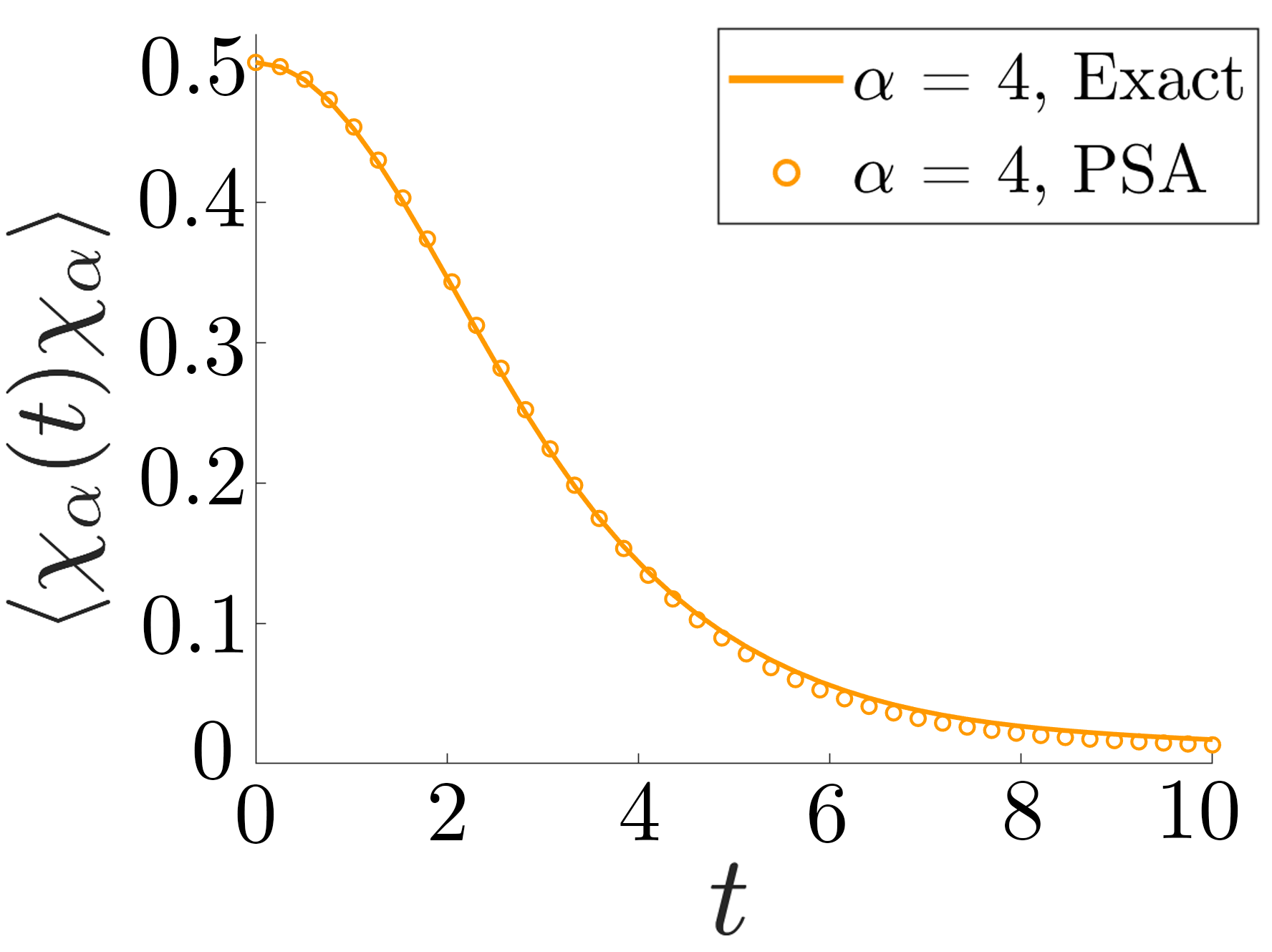}
  \label{fig:SYK_chi_4}
\end{subfigure}%
\begin{subfigure}{.245\textwidth}
  \centering
  \includegraphics[width=4.3cm]{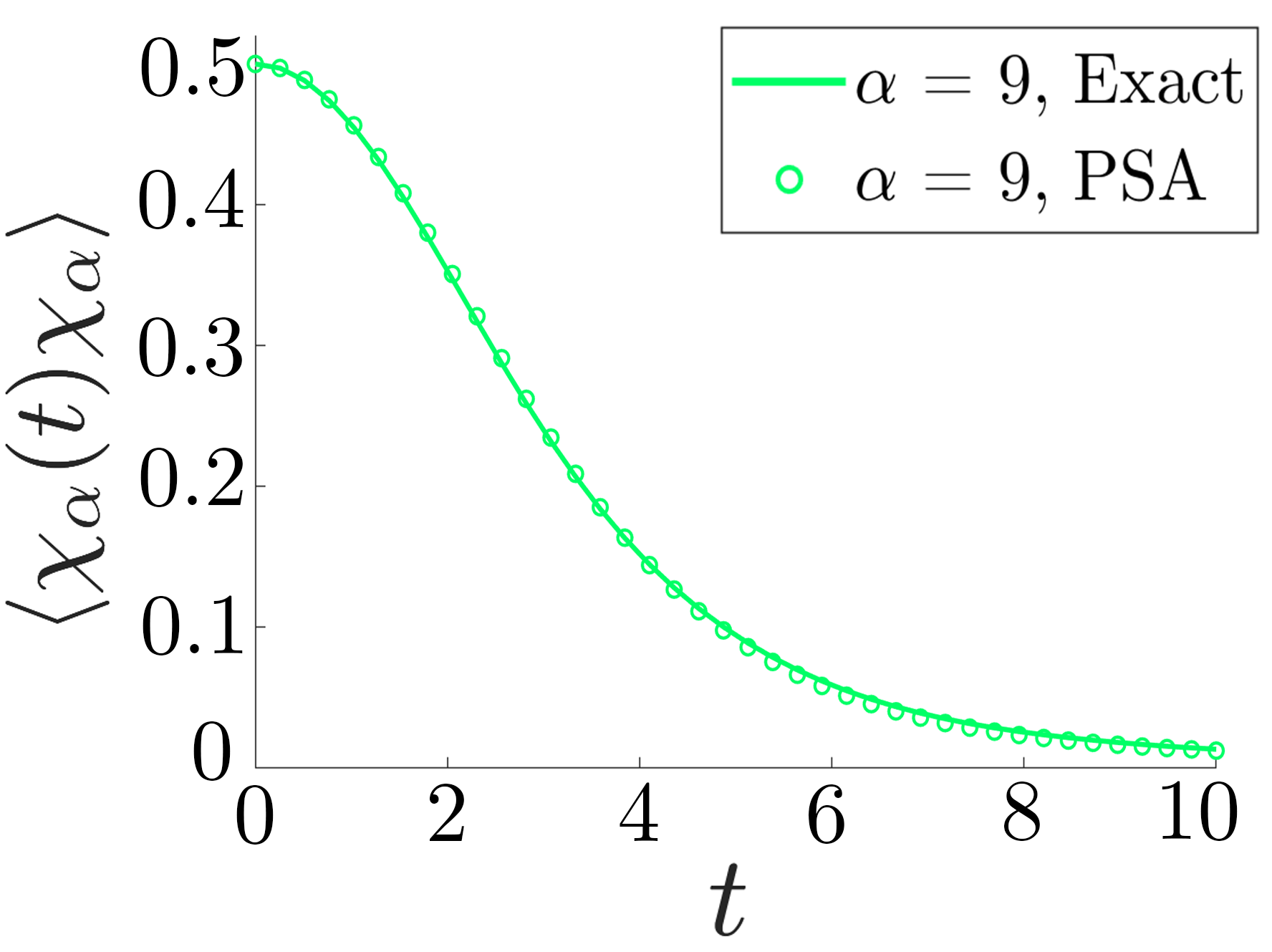}
  \label{fig:SYK_chi_9}
\end{subfigure}
\begin{subfigure}{.245\textwidth}
  \centering
  \includegraphics[width=4.3cm]{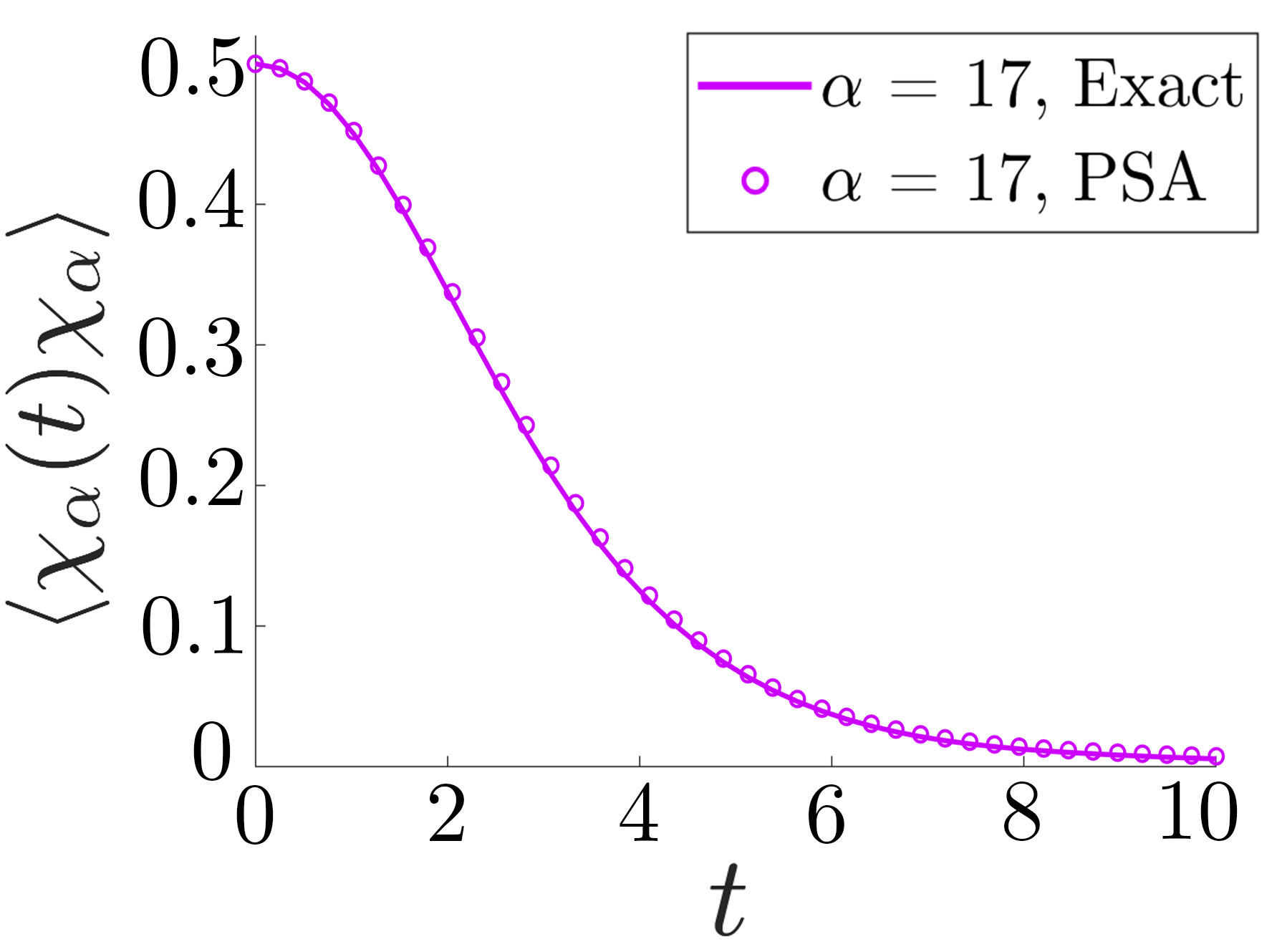}
  \label{fig:SYK_chi_17}
\end{subfigure}%
\begin{subfigure}{.245\textwidth}
  \centering
  \includegraphics[width=4.3cm]{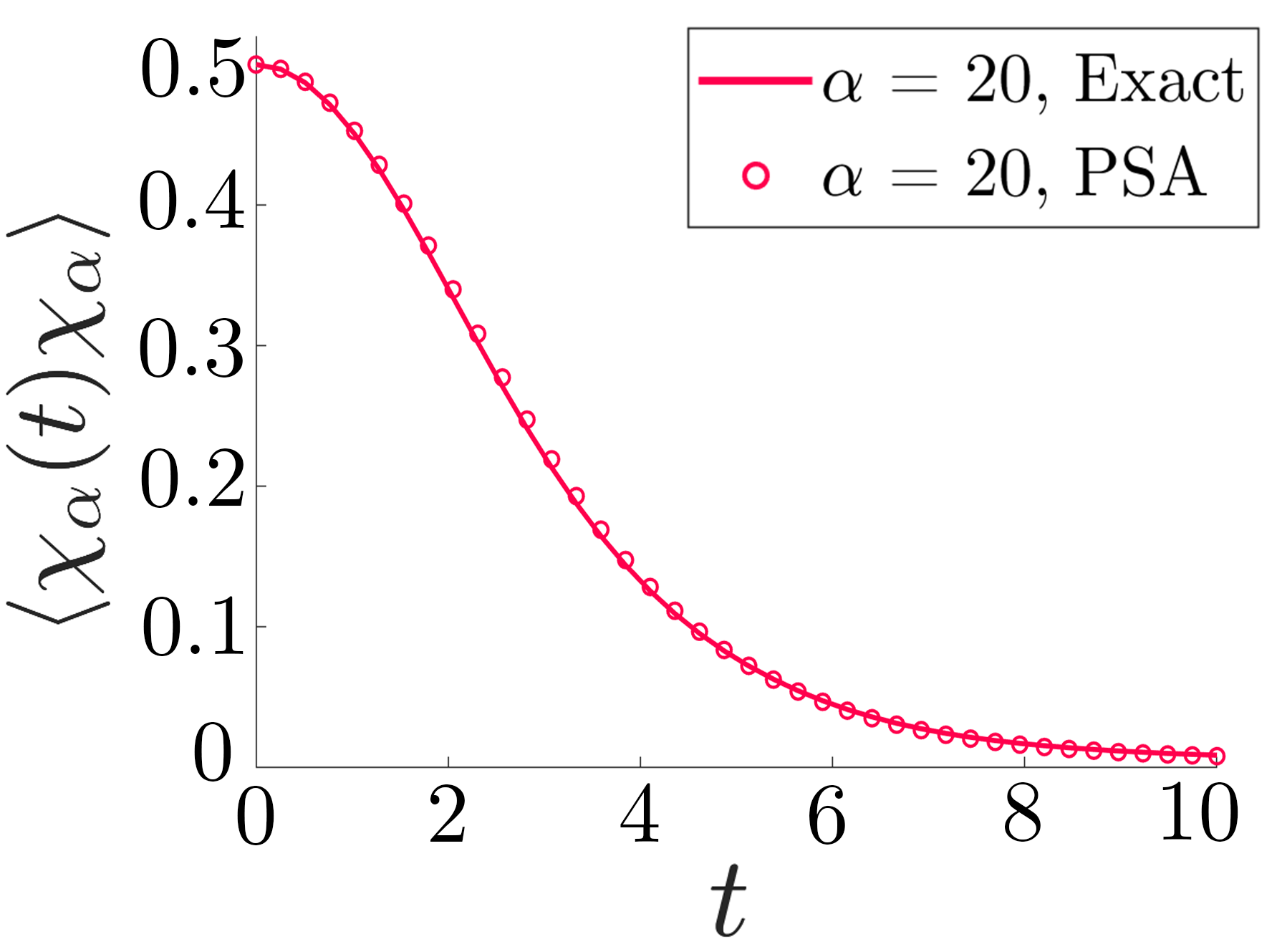}
  \label{fig:SYK_chi_20}
\end{subfigure}%
\caption{Exact and PSA correlation functions $\langle \chi_\alpha (t) \chi_\alpha \rangle$ for the SYK model with $N=20$, $T=0.7$, and $\ell=16$. $\alpha \in\{ 4,9,17,20\}$ are displayed as representatives of all the correlation functions.}
\label{fig:SYKCorrelation}
\end{figure}

To quantify the accuracy of the PSA applied to the SYK model, we study the entanglement entropy on subsystems of the thermal state, and local correlation functions. The particular quantities of interest to us are the subsystem entropy $S_i = S \big(\rho_i\big) = S\big( \text{tr}_{N/2-i}(\rho) \big)$ (where the partial trace runs over the last $N/2-i$ qubits), and the two-point correlation functions $\langle \chi_\alpha (t) \chi_\alpha \rangle$ for various $\alpha$. As a means for comparing the exact solution to the PSA, we consider the SYK model with $N=20$ Majorana fermions at $T=0.7$. For this scenario, we computed $S_i$ and  $\langle \chi_\alpha (t) \chi_\alpha \rangle$ for the exact thermal state and that produced by PSA with $\ell = 16$ layers of unitaries, using the exact Hamiltonian to implement time evolution. In general, more layers of unitary transformations were needed to capture the features of this nonlocal model than were needed for the Ising model. Figure \ref{fig:SYKEntropy} displays the subsystem entropies versus time, and Figure \ref{fig:SYKCorrelation} displays the correlation functions versus time.

As depicted in Figure \ref{fig:SYKEntropy}, $S_i$ calculated with the PSA agrees well with that of the exact solution across all subsystems, with all discrepancies less than $1.3 \%$. In addition, we see excellent agreement between the exact and PSA correlation functions at a variety of sites, as shown in Figure \ref{fig:SYKCorrelation}. Clearly then, given sufficiently many layers, the PSA is capable of capturing the essential features of the thermal state of a nonlocal Hamiltonian.

\subsection{Relation to Other Approaches} \label{sec:Relations}

The two defining features of the PSA that make it a tractable approximation are its ignorance of the exact spectrum and its locality.

The PSA requires no knowledge of the exact spectrum of the Hamiltonian. Given that calculating the exact spectrum for arbitrary systems is an exponentially difficult problem, this property is especially valuable, and allows the PSA to be extended to large and/or complex systems, where one cannot attain an analytic expression for the exact spectrum. Contrarily, some other approaches for approximately preparing thermal states require knowledge of the spectrum, such as that described in Reference \cite{pachos2018quantifying}. Due to the difficulty in obtaining such information on general many-body systems, these approaches are less applicable to large many-body systems than the PSA.

Furthermore, the locality of the PSA enables one to calculate expectation values of local observables with little overhead. For a PSA state with $\ell$ layers of unitaries, the calculation of a 1-site observable can be performed on a subsystem of $n' \leq 2 \ell$ qubits centered around the site of the observable, as demonstrated in Figure \ref{fig:LocalCalc}. As the subsystem size scales linearly with $\ell$, these calculations can be readily performed given a sufficiently small number of layers, which is the typical case for local models. Thus, it is quite straightforward to implement the PSA into systems described by local Hamiltonians, wherein the term $\mathrm{tr}(\rho_{\mathrm{PSA}} H)$ in $F$ can be computed locally on a classical computer. This property enables classical simulations of the PSA to be carried out on arbitrarily large many-body systems described by local interactions, granted that the number of layers is not excessively large. For such systems, the requisite calculations needed to implement the PSA are no more complex than they are for small systems.  As we will demonstrate in Section \ref{sec:LocalCalculations}, the PSA can accurately capture the behavior of large systems thanks to its locality.

An approach to thermal state preparation similar to the PSA is discussed in Reference \cite{kato2016quantum}, where it is proven that the thermal state of a 1D local Hamiltonian can be approximated by applying a depth-two quantum circuit to a product state, such that the unitary gates in the circuit act on $O\big(\log^2(n)\big)$ qubits. The structure of this approximation is akin to the PSA, but the approaches differ in that we use strictly two-qubit gates and a circuit of arbitrary depth $\ell$. In addition, the PSA is is not limited to local thermal states. Our benchmarking results for the SYK model and the results of Section \ref{sec:TFD} indicate that the PSA can be extended to the preparation of thermal states of non-local Hamiltonians and themofield double states.

Another approach to thermal state preparation is the ancilla method presented in Reference \cite{2011AnPhy.326...96S}. In this method, one calculates the thermal state at inverse temperature $\beta$ by performing imaginary time evolution on a purification of the infinite temperature thermal state. Explicitly, let us denote the purification of the infinite temperature thermal state by $| \psi_0 \rangle$. As it is a purification, $| \psi_0 \rangle$ exists on two copies of the system, the physical space and an ancillary space. The goal of this method is to prepare the state $| \psi_\beta \rangle = e^{-\beta H /2} | \psi_0 \rangle$, from which thermal averages can be computed as $\langle O \rangle_\beta = \frac{1}{\langle \psi_\beta | \psi_\beta \rangle}\langle \psi_\beta |O| \psi_\beta \rangle$. Typically, one uses a Trotter decomposition to approximate $e^{-\beta H/2}$ and attain an estimate of $|\psi_\beta \rangle$. When $H$ consists of only two body interactions, the Trotter decomposition of  $e^{-\beta H/2}$ has a spatial structure identical to the interspersed 2-qubit unitaries applied to the product mixed state of the PSA, as depicted in Figure \ref{fig:psa}. Hence, the ancilla method and the PSA share similar capabilities in the case of 2-body Hamiltonians. However, as we demonstrate with the SYK model, the PSA distinguishes itself from the ancilla method because it can be extended to the preparation of thermal states of non-local Hamiltonians, where the Trotter decomposition of $e^{-\beta H/2}$ does not take the form of interspersed 2-qubit transformations. In addition, the ancilla approach differs from the PSA in that it applies a predetermined non-unitary transformation to a fixed pure state defined on two copies of the system, which ultimately changes the spectrum of the density matrix. By contrast, the PSA applies a variable unitary transformation to a mixed state on the physical space, retaining the spectrum of the density matrix. We expect that the greater variability of the PSA makes it more versatile than the ancilla approach.

\section{Local calculations}\label{sec:LocalCalculations}
\begin{figure}[h]
  \centering
  \includegraphics[width=.48\textwidth]{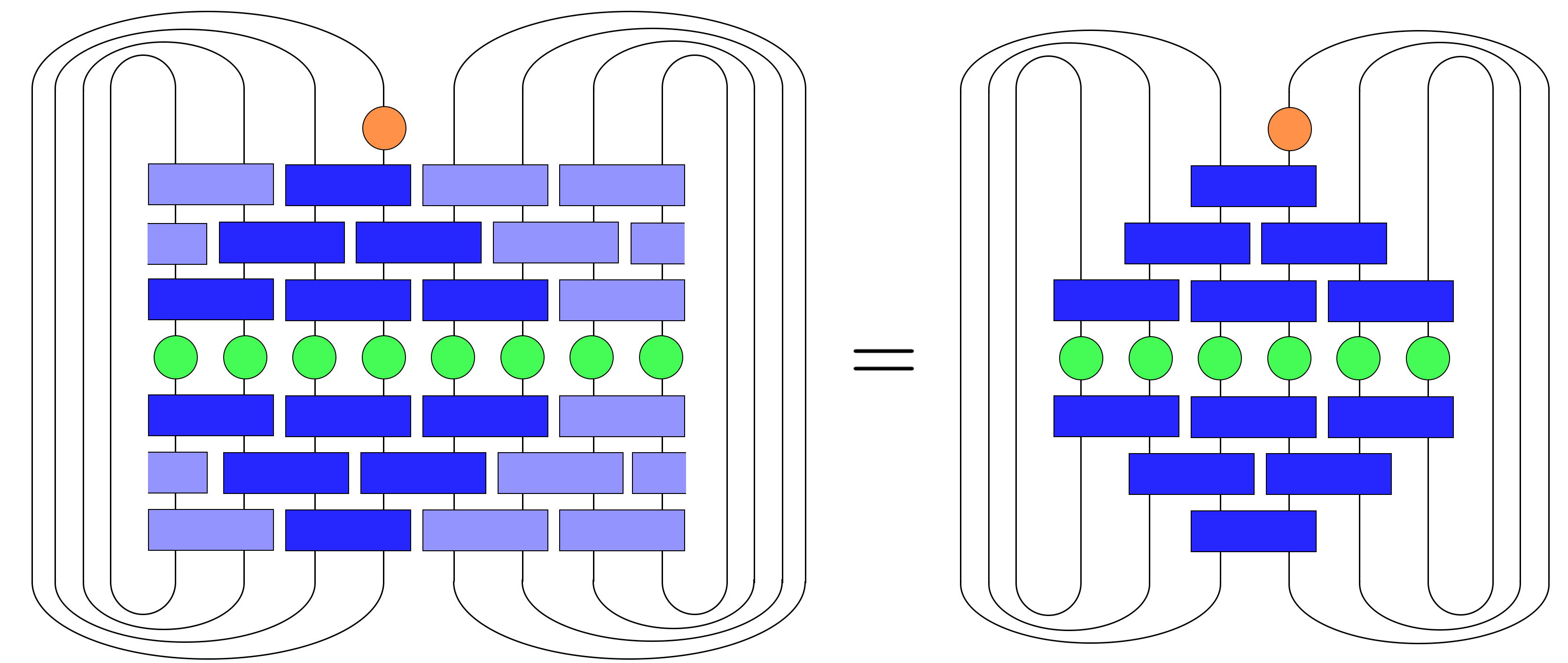}
  \caption{A tensor network depiction of a local calculation performed on a PSA state. This figure indicates that the expectation value of a 1-site observable on a PSA state with a circuit of depth $\ell=3$ can be performed on a subsystem of $n'=2\ell = 6$ qubits.} \label{fig:LocalCalc}
\end{figure}

As the PSA is constructed from local 2 qubit unitaries, it is intrinsically local. As we mentioned previously, this property allows the expectation values of local observables to be computed on subsystems around the site of the observable. Figure \ref{fig:LocalCalc} depicts a local calculation of this sort.

Using this property, one can apply the PSA to thermal states of large systems ($n \gg 1$) governed by local Hamiltonians without resorting to inefficient calculations on a $2^n$-dimensional Hilbert space. Accordingly, the variational calculations used to minimize the free energy of the PSA in classical simulations can be performed on small subsystems of the entire system, which may be arbitrarily large.

\begin{figure}[h]
\begin{subfigure}{.245\textwidth}
  \centering
  \includegraphics[width=4.6cm]{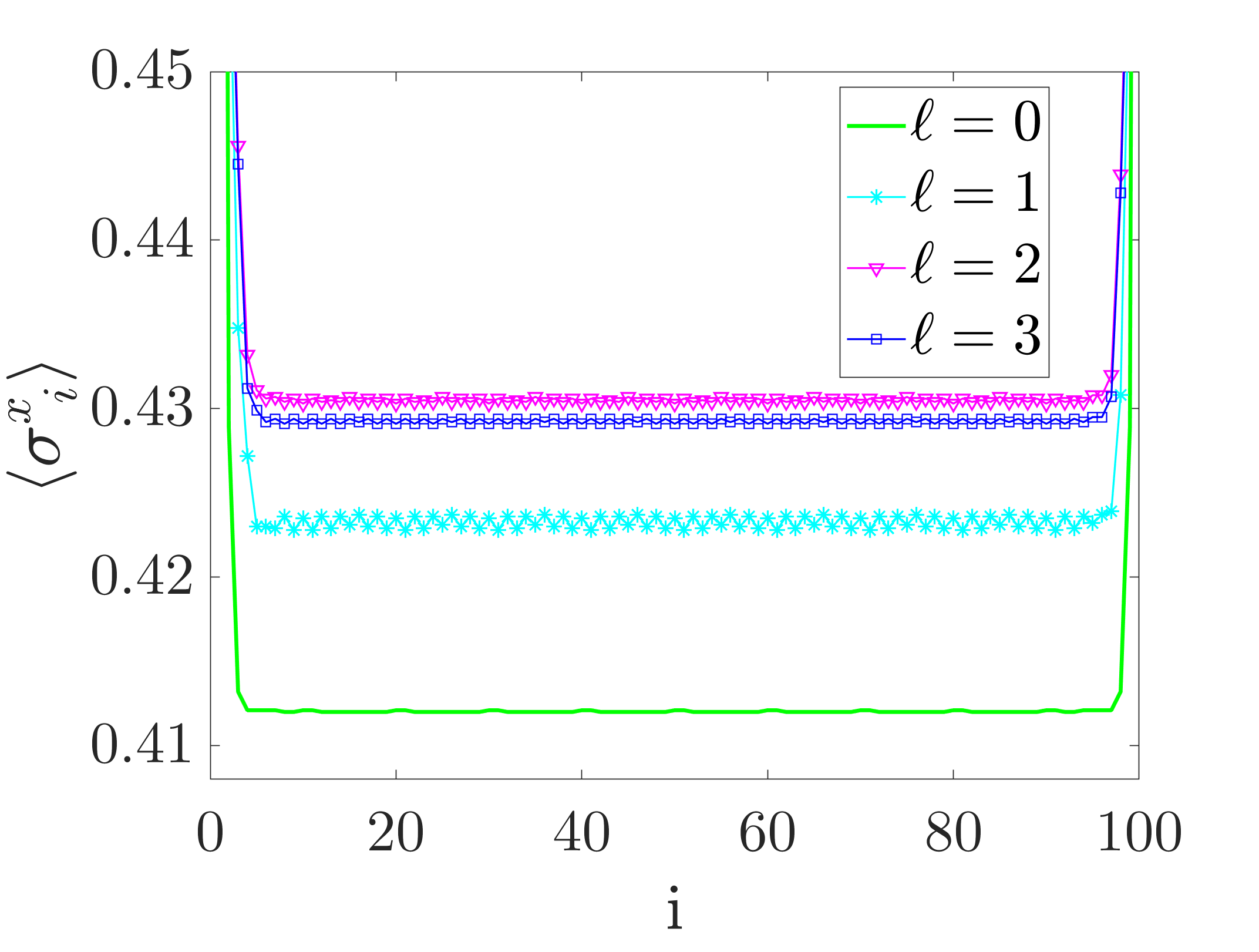}
  \label{fig:X_n=100}
\end{subfigure}
\begin{subfigure}{.245\textwidth}
  \centering
  \includegraphics[width=4.6cm]{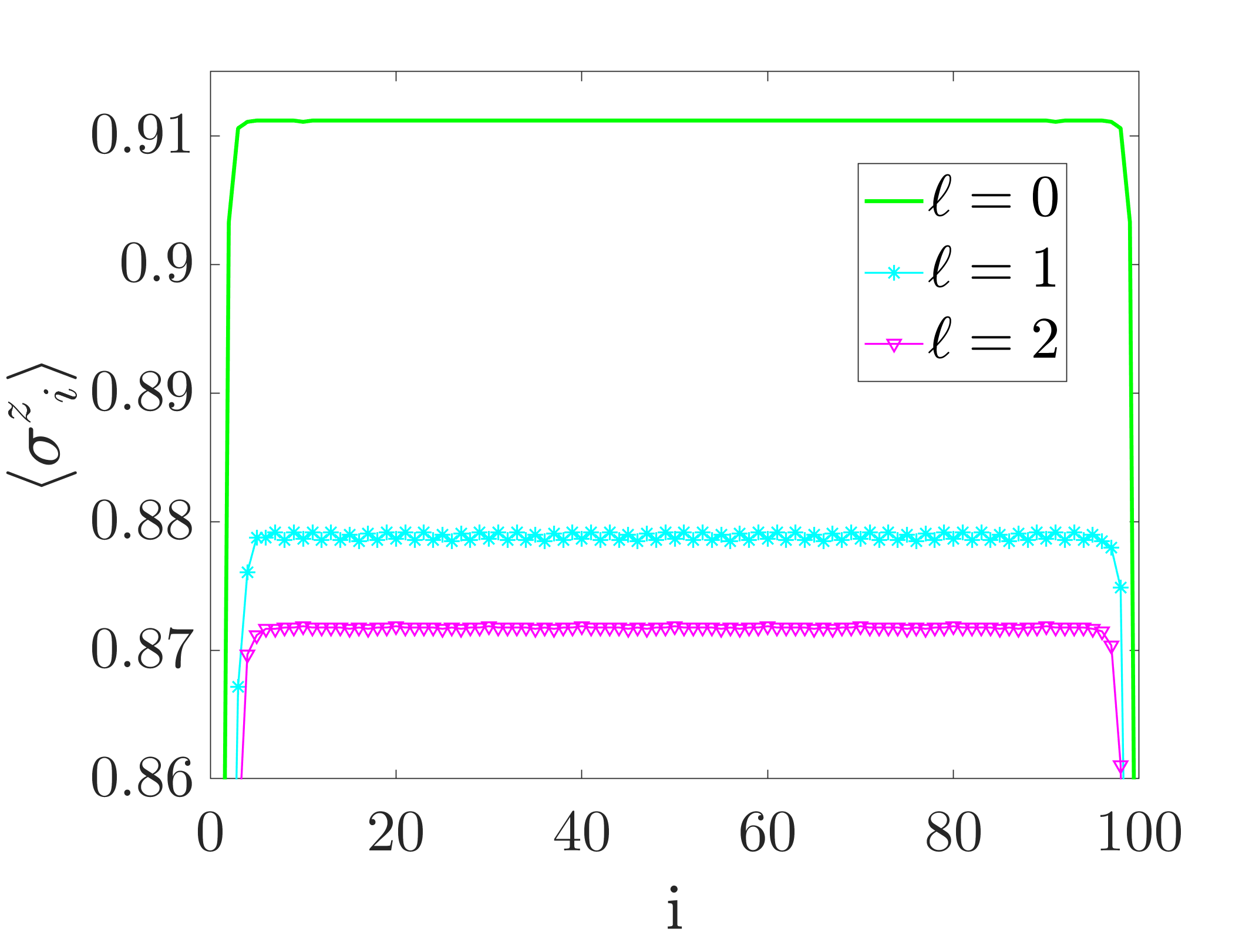}
  \label{fig:Z_n}
\end{subfigure}
\caption{Plots of $\langle \sigma^x_i \rangle $ (\textbf{left}) and $\langle \sigma^z_i \rangle $ (\textbf{right}) for a PSA state of the mixed field Ising chain with $n=100$, $h_x = 1.05$, and $ h_z = 0.5$ at $T=1.0$.}
\label{fig:LargeSystem}
\end{figure}

We applied this procedure to the mixed field Ising chain with $n=100$ spins and couplings ($h_x = 1.05$, $h_z = 0.5$). To study the behavior of the PSA on this system, we calculated the expectation values of one site observables, $\langle \sigma^x_i \rangle$ and $\langle \sigma^z_i \rangle$, on the PSA state. Our results are displayed in Figure \ref{fig:LargeSystem}.

Figure \ref{fig:LargeSystem} demonstrates that the PSA thermal state behaves as expected for the mixed field Ising chain at large $n$. Away from the boundaries (where we have employed open boundary conditions), the Hamiltonian is essentially translation invariant, so one would naturally expect $\langle \sigma^x_i \rangle$ and $\langle \sigma^z_i \rangle$  to take constant values throughout the interior of the chain. This indeed agrees with our results, wherein $\langle \sigma^x_i \rangle$ and $\langle \sigma^z_i \rangle$  are uniform away from the boundaries. In addition, we note that the expectation values of these one site observables begin to converge to stable values after just a few layers are applied to the product state. This fast convergence indicates that the PSA state quickly reaches its optimal thermal state approximation for local models.

\section{Approximate preparation of thermofield double states}\label{sec:TFD}

In addition to the thermal state, it is also of interest to prepare the thermofield double (TFD) state of a many-body system. The TFD state is a purification of the thermal state, appearing in the basis of energy eigenstates, $|E_i\rangle $, as
\begin{equation}
\ket{\mathrm{TFD}} = \sum_i \sqrt{\frac{e^{-\beta E_i}}{Z}} \ket{E_i}_L\ket{E_i}_R,
\end{equation}
where the subscripts $L$ and $R$ denote two identical copies of the system. As a purification, its defining property is that its partial trace over system $R$ is the thermal state:
\begin{equation}
\mathrm{tr}_R \big(\ket{\mathrm{TFD}} \bra{\mathrm{TFD}}\big) = \frac{e^{-\beta H}}{Z} = \rho.
\end{equation}

The procedure to prepare the approximate TFD state has two steps. First, one prepares a symmetric purification of $\rho_{\mathrm{prod}}$ which consists of pairs of entangled spins with variable degrees of entanglement related to the $\{\epsilon_r\}$. Then one applies the PSA unitary $U$ to both sides of the resulting purification to produce an approximation to the TFD state.

To probe the quality of the approximation, it is instructive to examine the expectation values of operators of the form $O = O_L \otimes O_R$ in the TFD state. This quantity can be expressed as
\begin{equation}\label{eq:expectation}
\langle \mathrm{TFD} | O_L \otimes O_R | \mathrm{TFD} \rangle = \text{tr} \big( \sqrt{\rho} O_L \sqrt{\rho} O_R^{\mathbb{T}} \big).
\end{equation}
Here $\mathbb{T}$ denotes transpose in the energy basis. For example, connected correlation functions fall under this class of operators:
\begin{equation}
\begin{split}
\langle O_L \otimes O_R \rangle_{\text{connected}} = \quad \qquad \\ \langle \mathrm{TFD} | O_L \otimes O_R | \mathrm{TFD} \rangle - \langle O_L \rangle \langle O_R \rangle = \\
\text{tr}\big( \sqrt{\rho} O_L \sqrt{\rho} O_R^{\mathbb{T}} \big) - \text{tr} \big( \rho O_L \big)\text{tr} \big( \rho O_R \big)
\end{split}
\end{equation}

Using the relation in Equation \ref{eq:expectation}, we can use the PSA to approximate the expectation values of these operators by replacing $\rho$ with $\rho_{PSA}$. These correlation functions partially characterize the TFD state, enabling us to test the PSA approximation of the TFD without explicitly constructing the purification.

\subsection{Mixed Field Ising Model}

We now seek to establish the accuracy of the PSA in preparing TFD states of the mixed field Ising model. We study the expectation values of the Pauli operators, $\langle \sigma^x_i \otimes \sigma^x_i \rangle$ and $\langle \sigma^z_i \otimes \sigma^z_i \rangle$,  and their corresponding connected correlation functions, $\langle \sigma^x_i \otimes \sigma^x_i \rangle_C :=  \langle \sigma^x_i \otimes \sigma^x_i \rangle - \langle \sigma^x_i \rangle \langle \sigma^x_i \rangle$  and $\langle \sigma^z_i \otimes \sigma^z_i \rangle_C :=  \langle \sigma^z_i \otimes \sigma^z_i \rangle - \langle \sigma^z_i \rangle \langle \sigma^z_i \rangle$. These quantities describe the interactions among the spins in the $L$ and $R$ subsystems of TFD state.

\begin{figure}[h]
\begin{subfigure}{.24\textwidth}
  \centering
  \includegraphics[width=4.6cm]{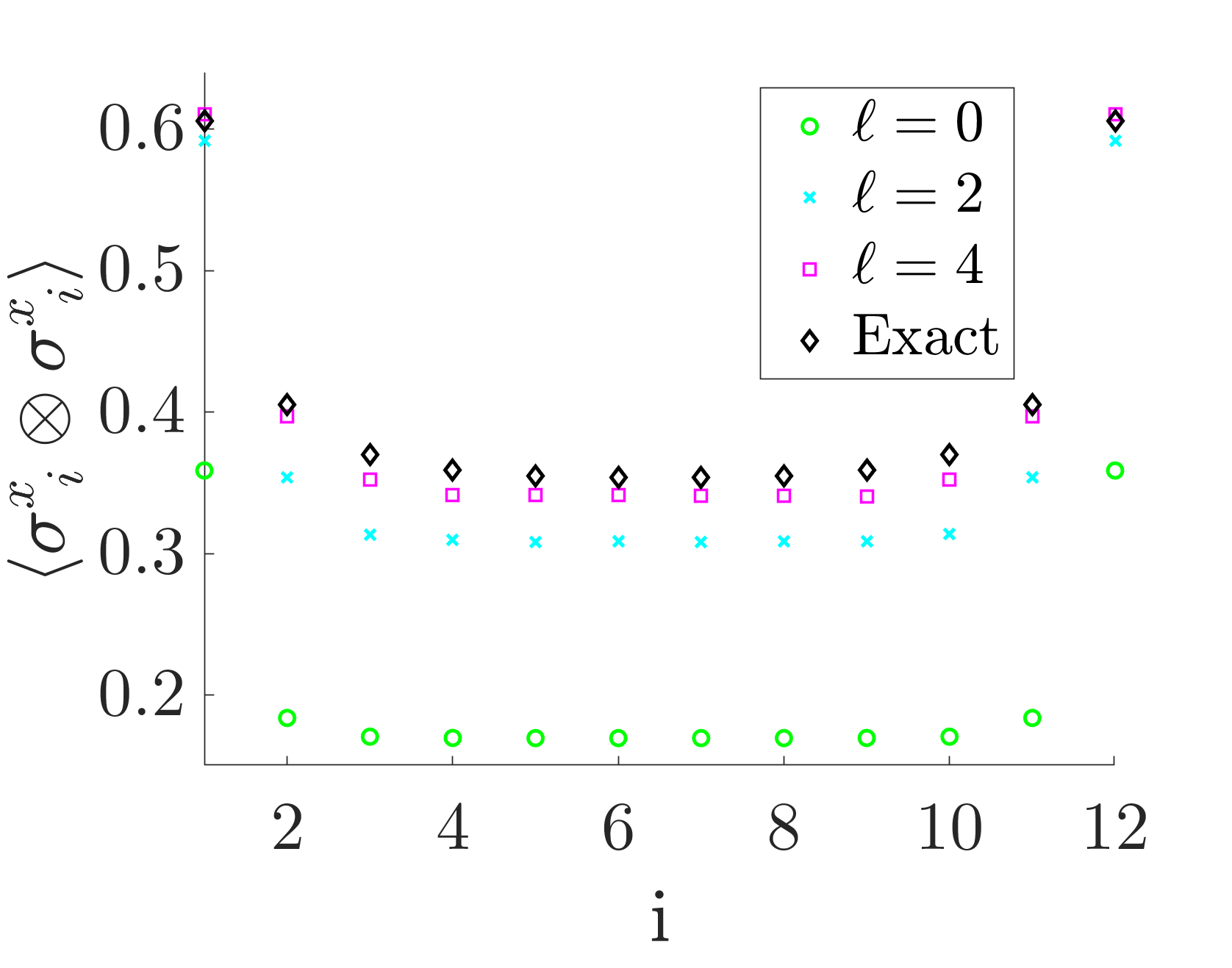}
  \label{fig:XX_n=100}
\end{subfigure}
\begin{subfigure}{.24\textwidth}
  \centering
  \includegraphics[width=4.6cm]{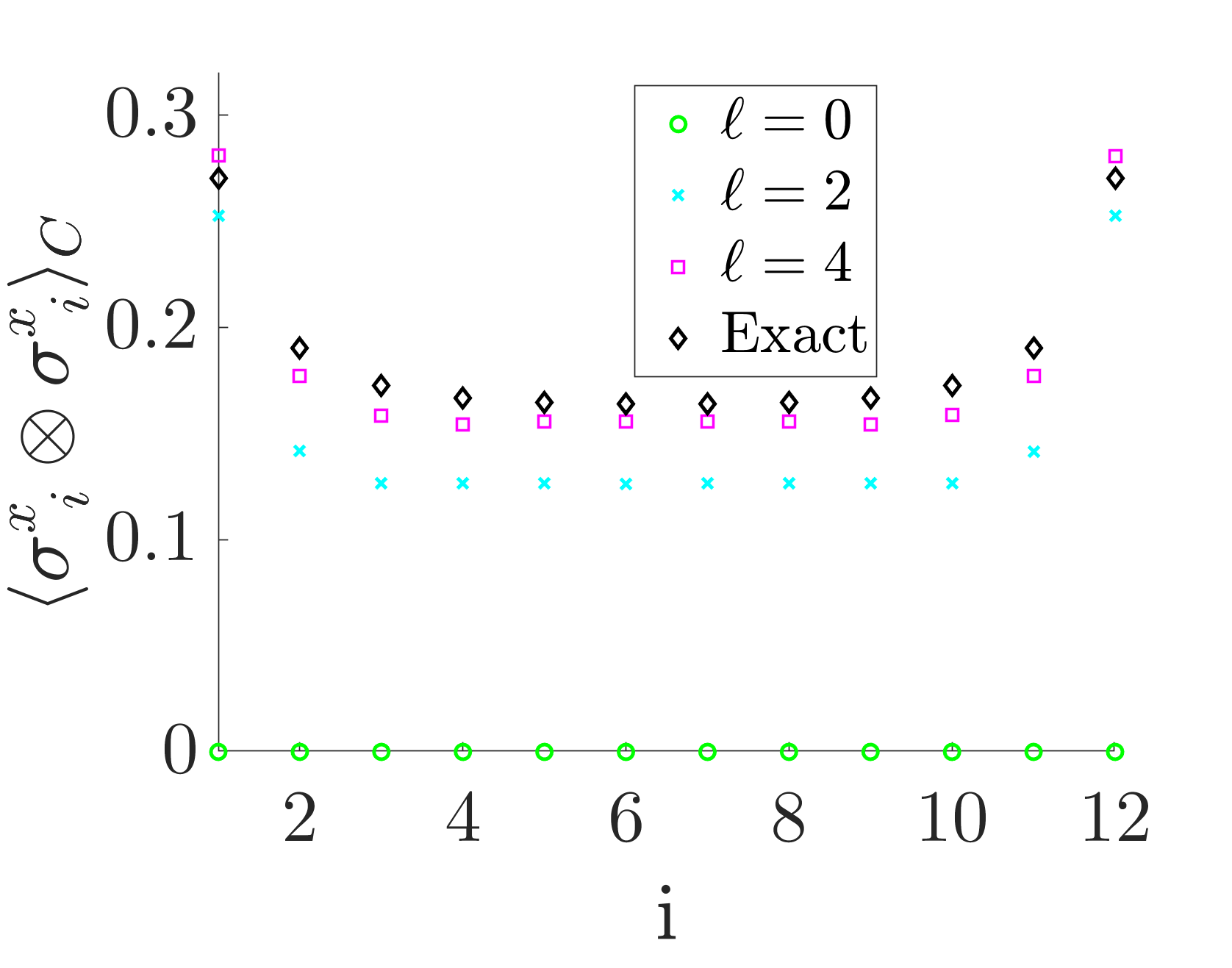}
  \label{fig:XX_c_n}
\end{subfigure}
\begin{subfigure}{.24\textwidth}
  \centering
  \includegraphics[width=4.6cm]{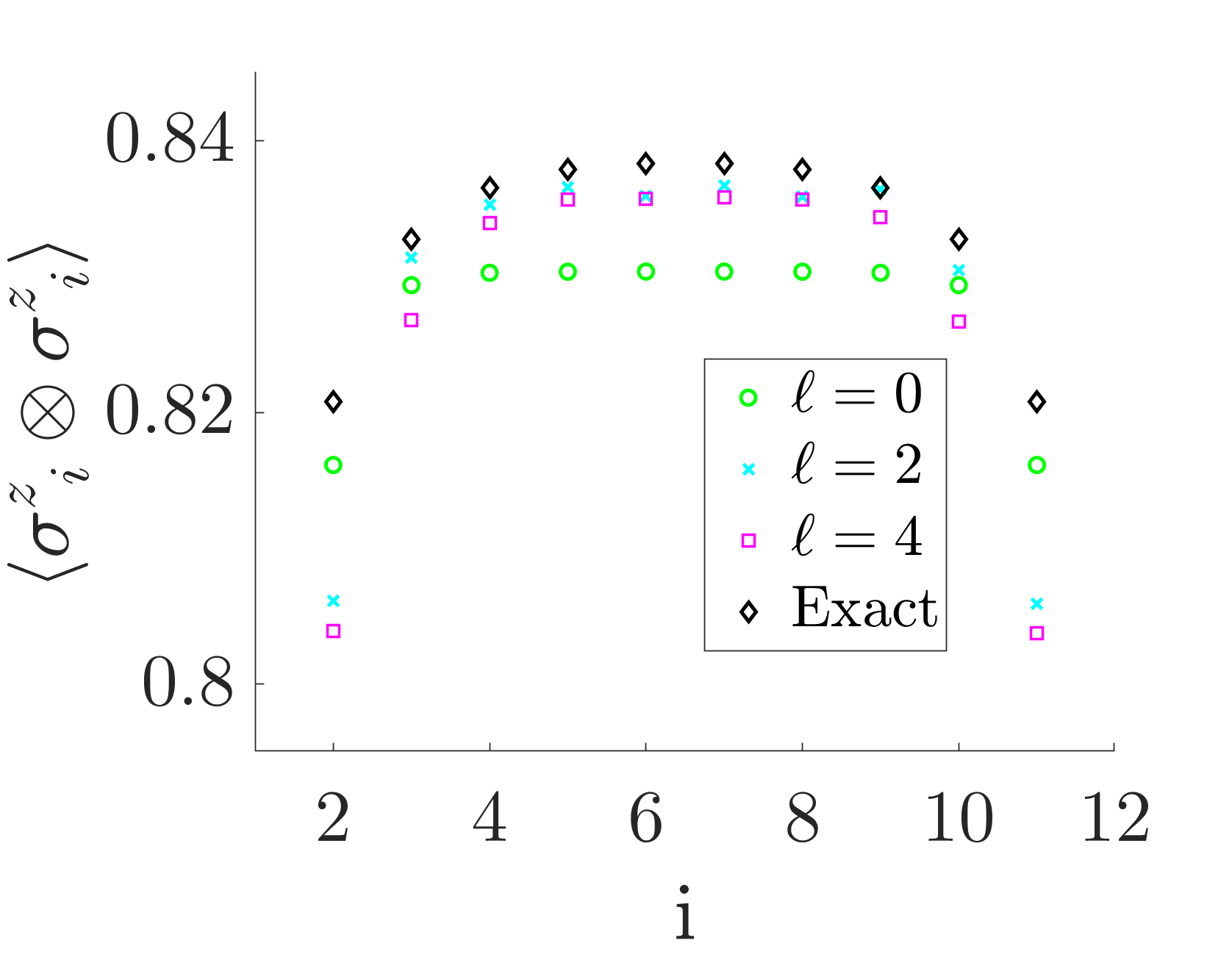}
  \label{fig:ZZ_n=100}
\end{subfigure}
\begin{subfigure}{.24\textwidth}
  \centering
  \includegraphics[width=4.6cm]{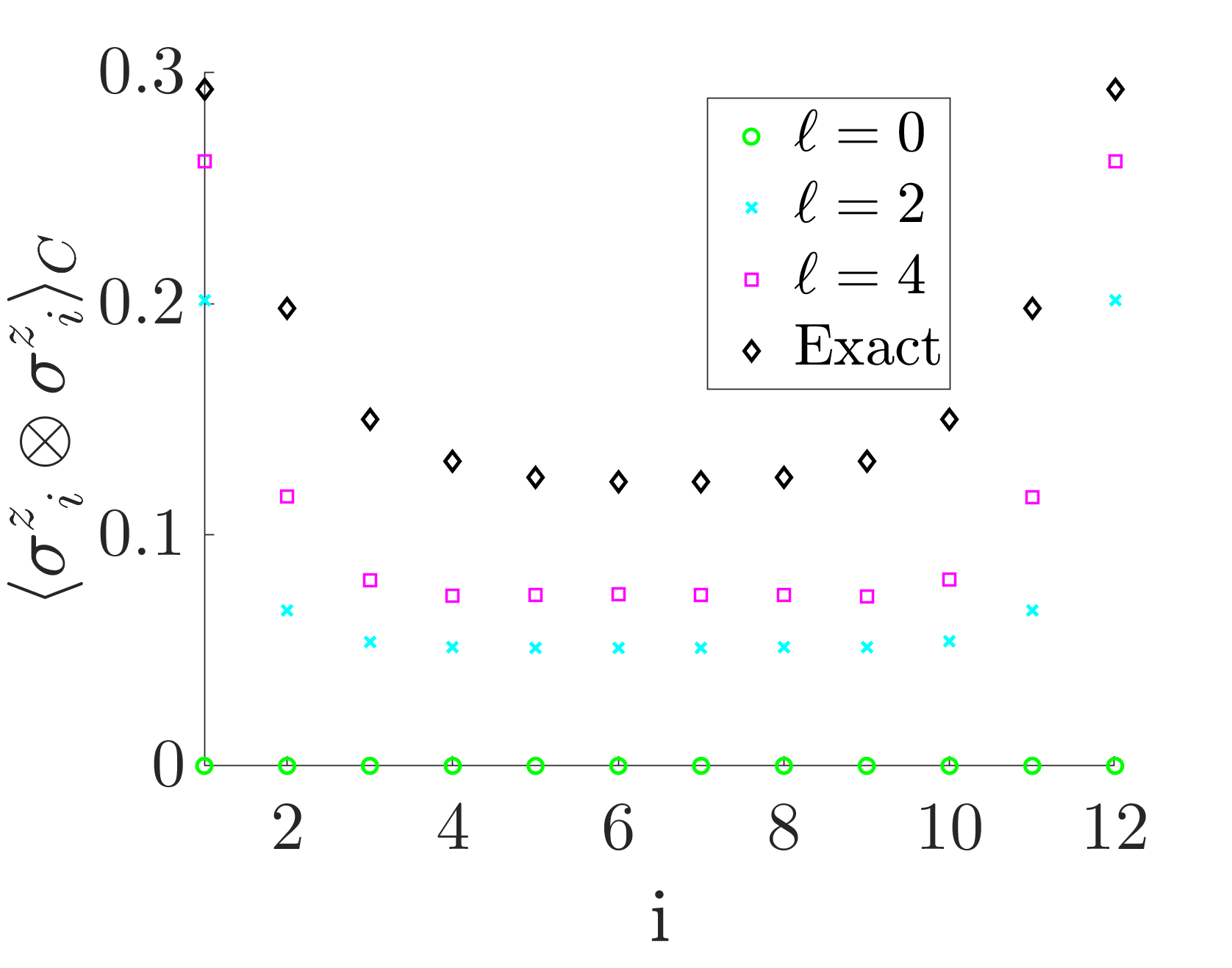}
  \label{fig:ZZ_c_n=100}
\end{subfigure}%
\caption{TFD expectation values: plots of $\langle \sigma^x_i \otimes \sigma^x_i \rangle $ (\textbf{upper left}), $\langle \sigma^x_i \otimes \sigma^x_i \rangle_C$ (\textbf{upper right}), $\langle \sigma^z_i \otimes \sigma^z_i \rangle $ (\textbf{lower left}), and $\langle \sigma^z_i \otimes \sigma^z_i \rangle_C $ (\textbf{lower right}) in the mixed field Ising TFD with $n=12,\  h_x = 1.05,$ and $h_z = 0.5$ at $T=1.0$. Exact results and PSA results at different layer numbers ($\ell$) are displayed. Here we note that the PSA has been generalized to allow the uncorrelated state $\rho_{\mathrm{prod}}$ to be diagonal in an arbitrary local basis.}
\label{fig:IsingCorrelation}
\end{figure}

We probe each of these quantities for a small system of $n=12$ spins, such that we can compare our approximate results with those of the exact solution. Again, we analyze the nonintegrable model with couplings ($h_x = 1.05, h_z  = 0.5$) at $T=1.0$. Our results are displayed in Figure \ref{fig:IsingCorrelation}.

From Figure \ref{fig:IsingCorrelation}, we see that TFD expectation values computed with the PSA agree well with the exact values, converging after just a few layers. Most notably, the plots of the connected correlation functions demonstrate how the PSA builds correlations as the number of layers increase. With $0$ layers ($\ell = 0$), the PSA is a product state, and the connected correlation functions vanish. As the number of layers increase, the unitary gates applied to the product state enable correlations and entanglement among the spins, and the approximate connected correlation functions become nonzero and approach their exact values. These positive results verify our claim that the PSA can be extended to the preparation of TFD states.

\subsection{SYK Model}
\subsubsection{Product Spectrum Approximation}
\begin{figure}[h]
\begin{subfigure}{.24\textwidth}
  \centering
  \includegraphics[width=4.25cm]{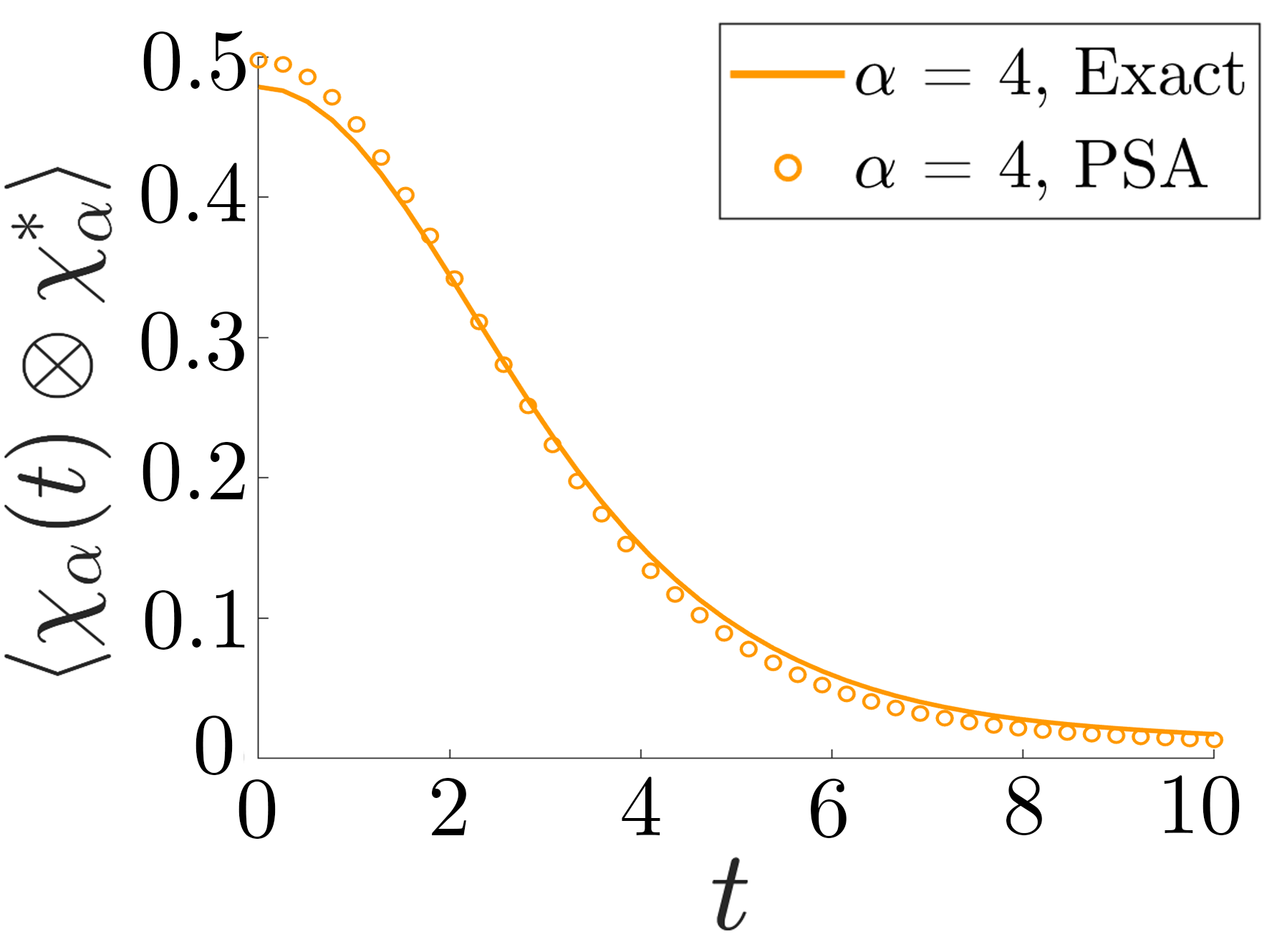}
  \label{fig:SYK_TFD_4}
\end{subfigure}%
\begin{subfigure}{.24\textwidth}
  \centering
  \includegraphics[width=4.25cm]{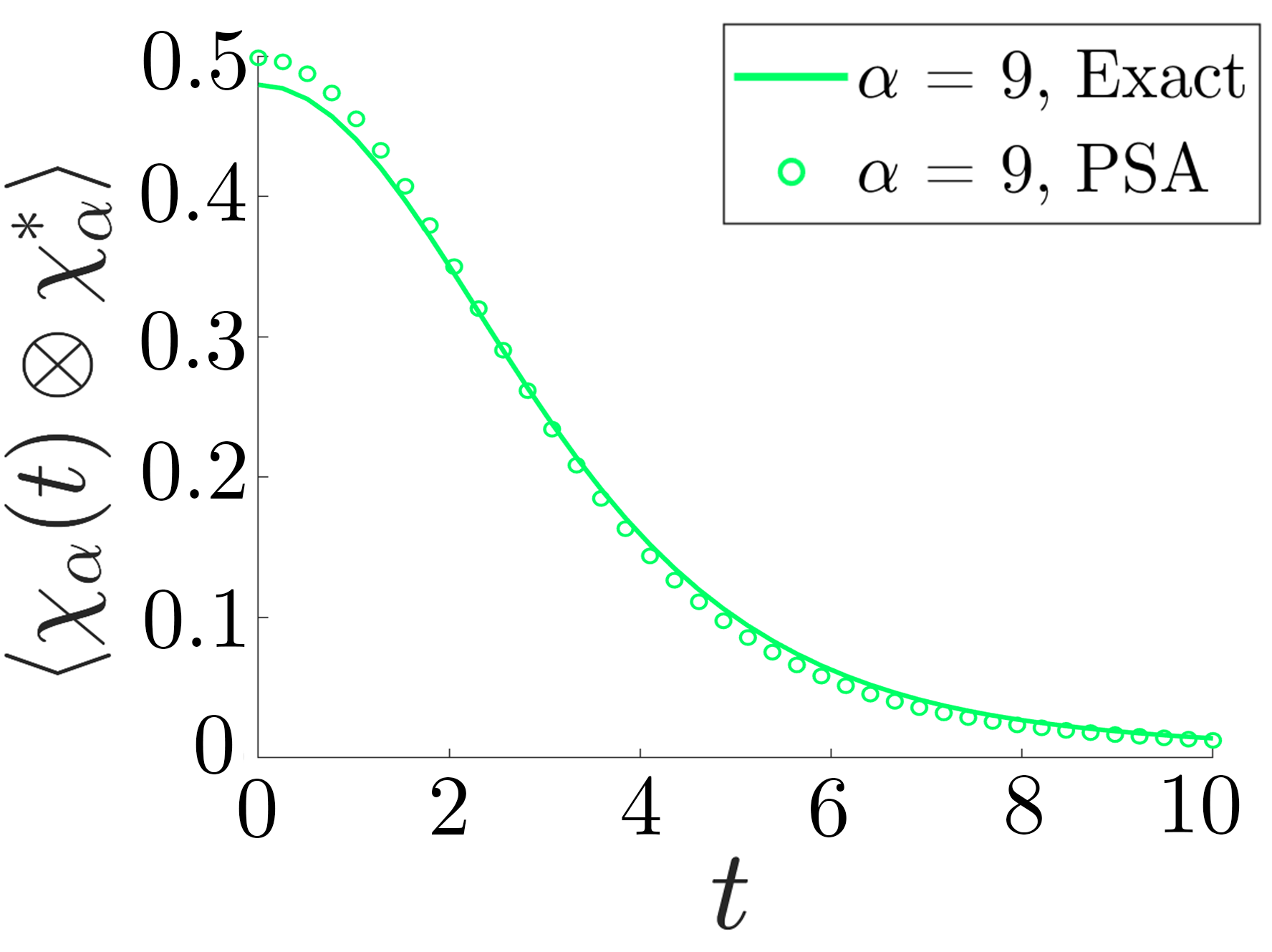}
  \label{fig:SYK_TFD_9}
\end{subfigure}
\begin{subfigure}{.24\textwidth}
  \centering
  \includegraphics[width=4.25cm]{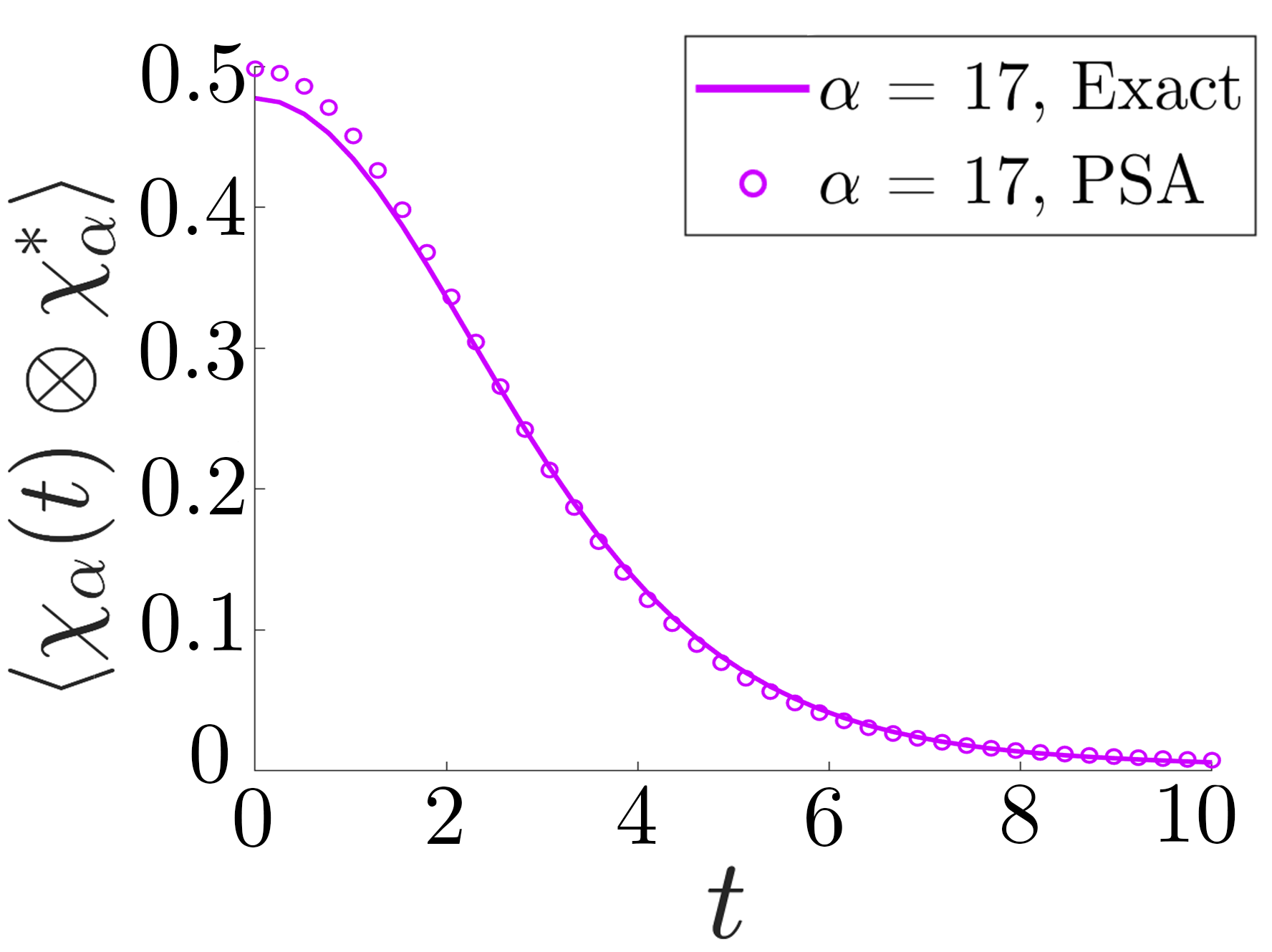}
  \label{fig:SYK_TFD_17}
\end{subfigure}%
\begin{subfigure}{.24\textwidth}
  \centering
  \includegraphics[width=4.25cm]{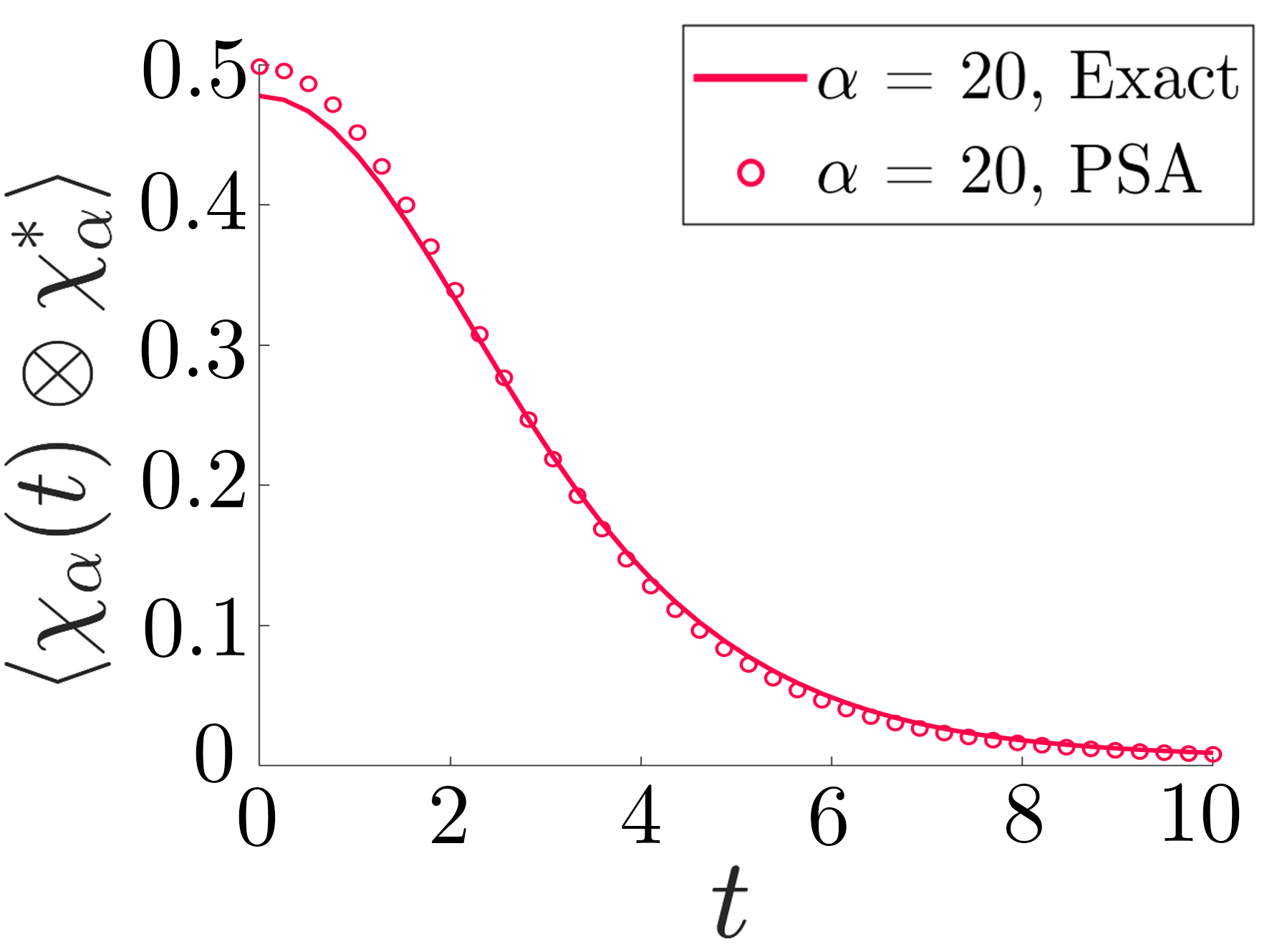}
  \label{fig:SYK_TFD_20}
\end{subfigure}%
\caption{Exact and PSA TFD correlation functions $\langle \chi_\alpha (t) \otimes \chi_\alpha^* \rangle$ for the SYK model with $N=20$ Majorana fermions per subsystem, $T=0.7$, and $\ell=16$. $\alpha \in\{ 4,9,17,20\}$ are displayed as representatives of all correlation functions.}
\label{fig:SYKTFDCorrelation}
\end{figure}

We now apply the PSA to the preparation of TFD states of the SYK model. Using the exact TFD state and the PSA TFD state, we compute the correlation functions $\langle \chi_\alpha(t) \otimes \chi_\alpha^* \rangle$ for various $\alpha$, using the exact Hamiltonian for time evolution. Such a quantity describes the correlations among Majorana fermions in the $L$ and $R$ subsystems at various sites. Figure \ref{fig:SYKTFDCorrelation} displays a comparison of the exact and PSA results with $N=20$ Majorana fermions per subsystem at $T=0.7$.

Figure \ref{fig:SYKTFDCorrelation} indicates good agreement between the exact and approximate TFD correlation functions for a variety of $\alpha$ over the time range $t\in[0,10]$. We see slight deviations between these quantities at early time, and the best agreement at late times. Nevertheless, the PSA accurately captures the dynamics of the TFD state of the SYK model.

\begin{figure}[h]
\begin{subfigure}{.24\textwidth}
  \centering
  \includegraphics[width=4.25cm]{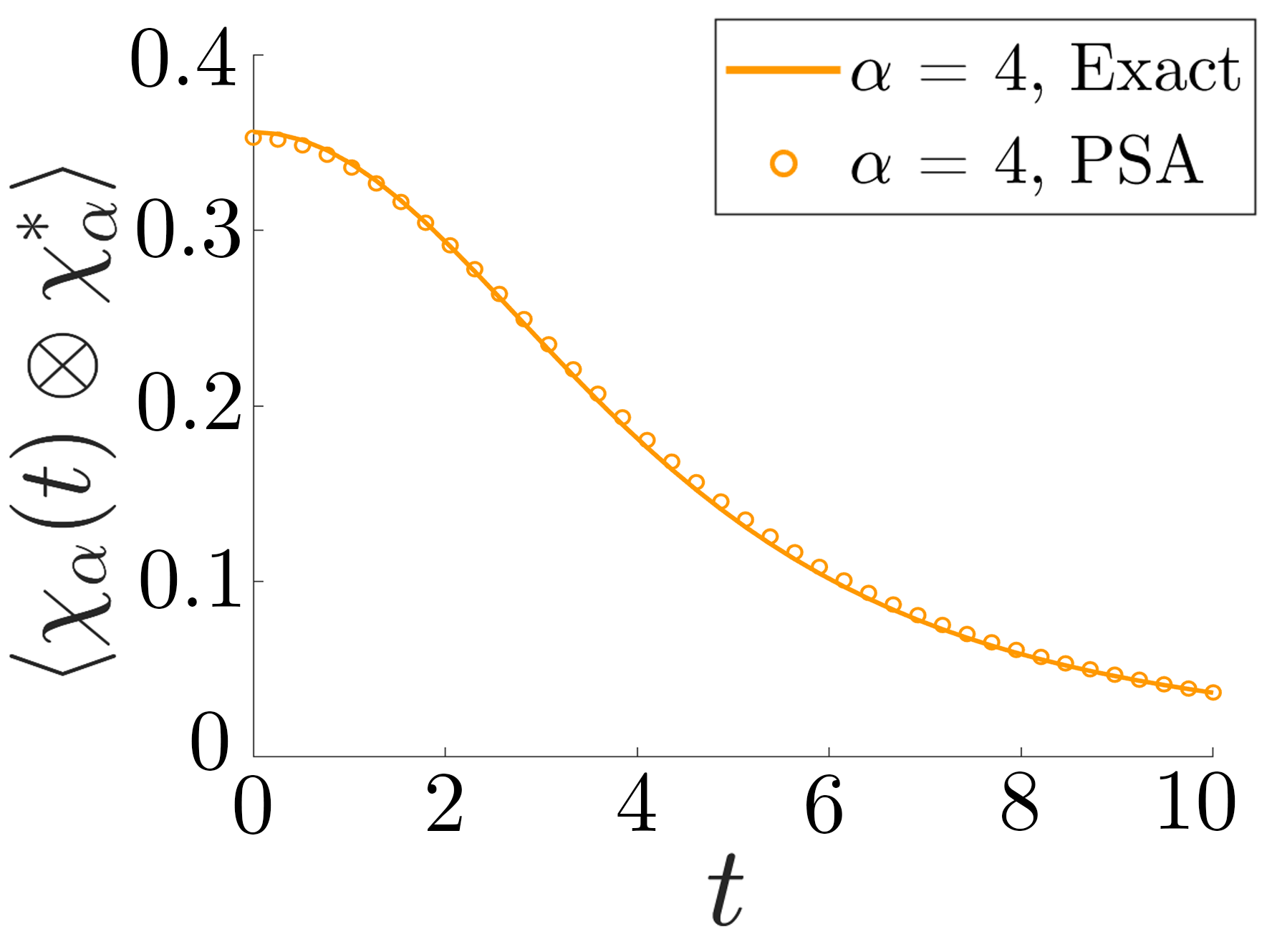}
  \label{fig:SYK2_TFD_4}
\end{subfigure}%
\begin{subfigure}{.24\textwidth}
  \centering
  \includegraphics[width=4.25cm]{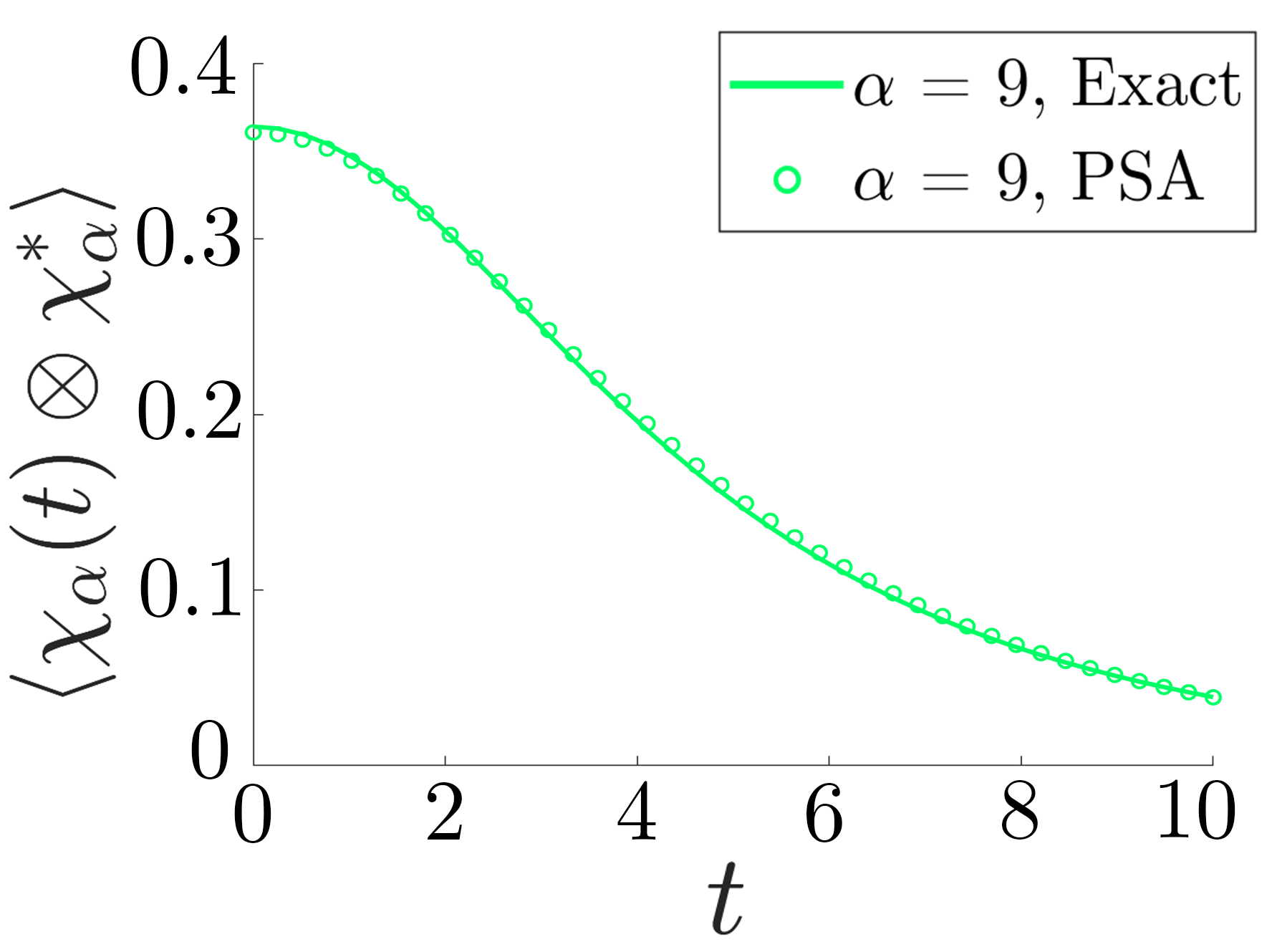}
  \label{fig:SYK2_TFD_9}
\end{subfigure}
\begin{subfigure}{.24\textwidth}
  \centering
  \includegraphics[width=4.25cm]{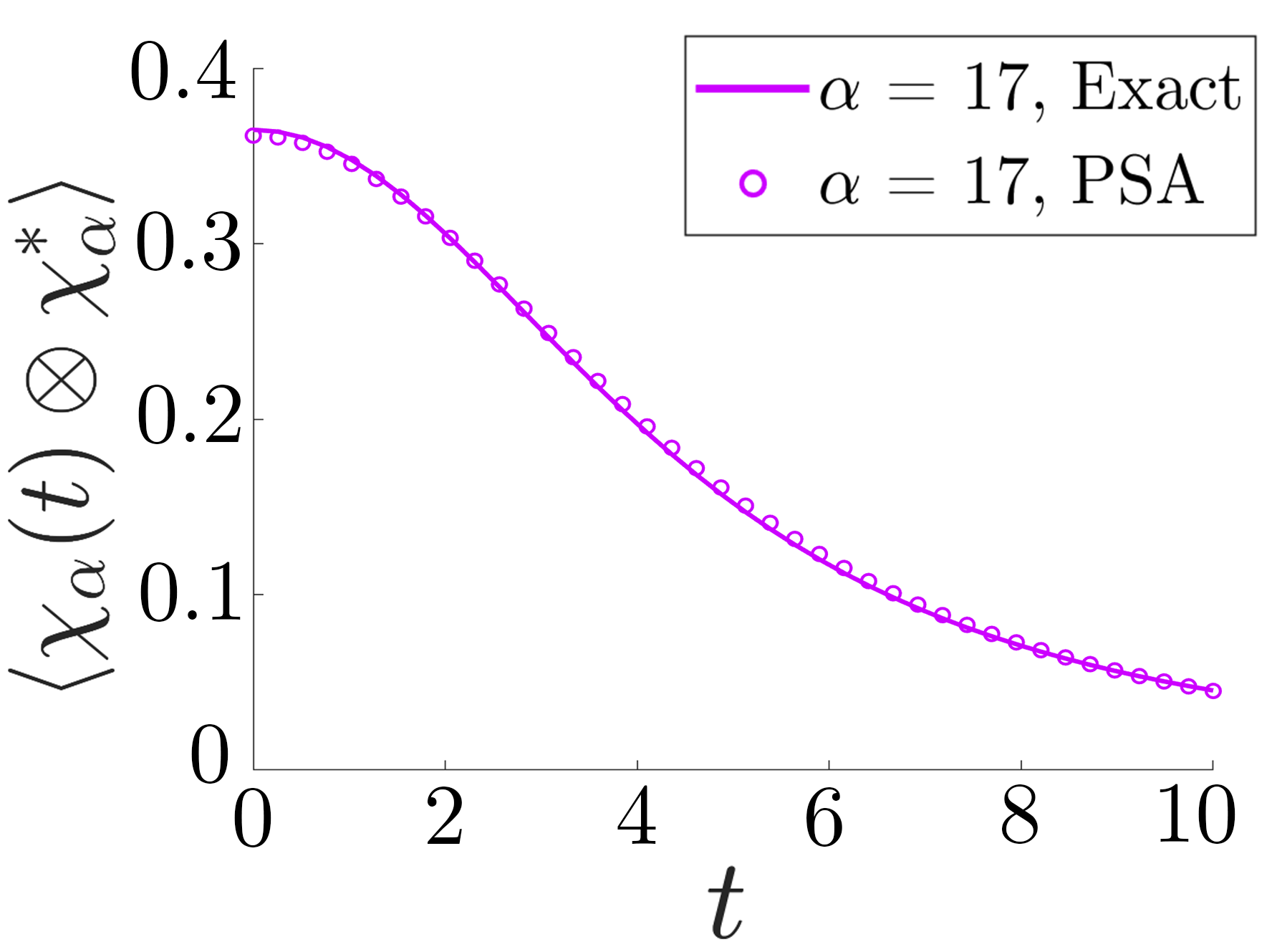}
  \label{fig:SYK2_TFD_17}
\end{subfigure}%
\begin{subfigure}{.24\textwidth}
  \centering
  \includegraphics[width=4.25cm]{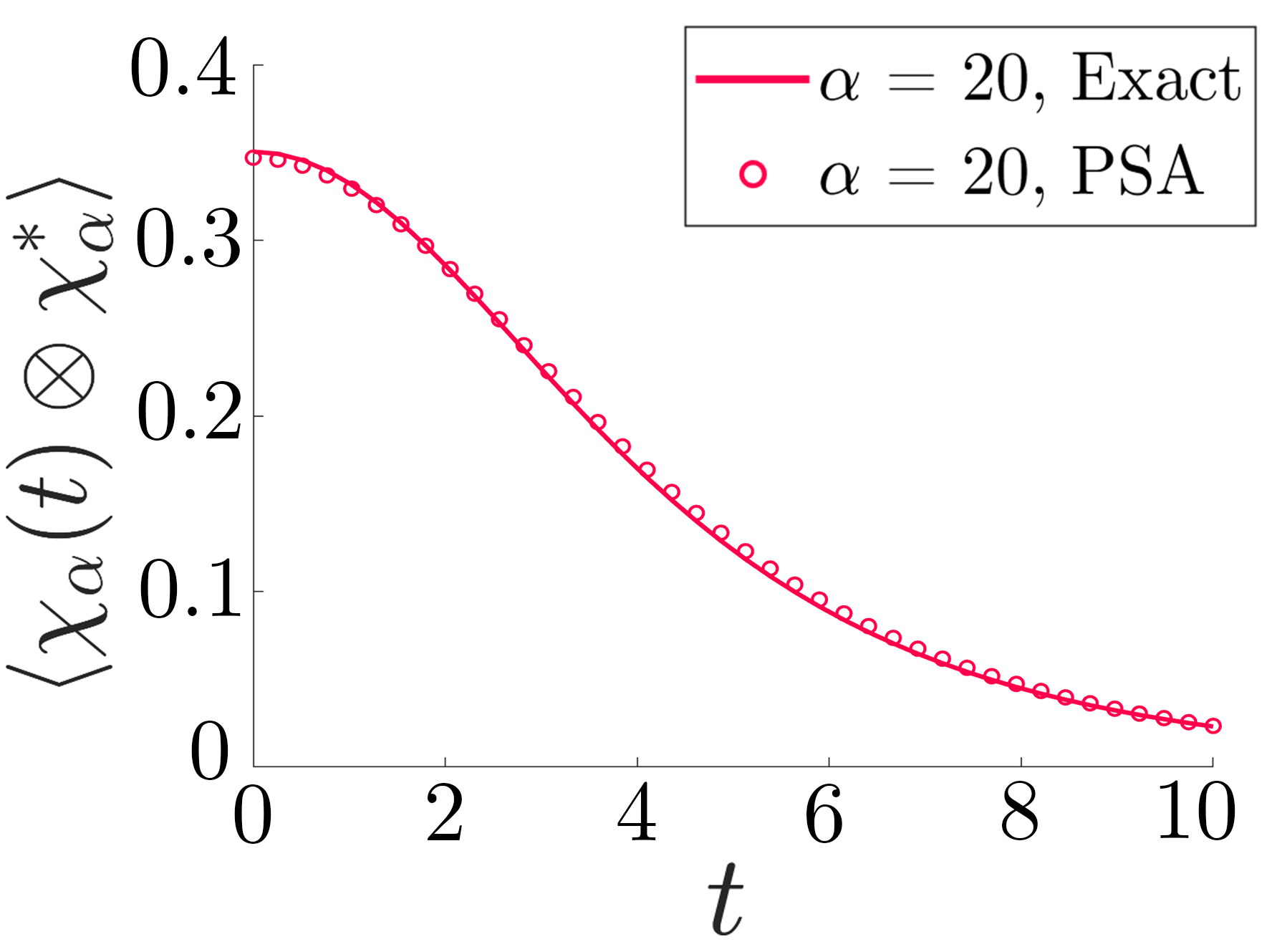}
  \label{fig:SYK2_TFD_20}
\end{subfigure}
\caption{Exact and PSA (using exact unitary) TFD correlation functions $\langle \chi_\alpha (t) \otimes \chi_\alpha^* \rangle$ for the SYK model with $N=20$ Majorana fermions per subsystem and $T=0.2$. $\alpha \in\{ 4,9,17,20\}$ are displayed as representatives of all correlation functions.}
\label{fig:SYK2TFDCorrelation}
\end{figure}

In extending the PSA to the SYK model at lower temperatures, say $T=.2$ and below, we found that the PSA state obtained from minimizing the free energy struggled to capture some physical properties. While correlations in the mixed thermal state were reproduced reasonably well, correlations in the TFD state were not well captured. To determine if the problem is with the PSA itself, or with the circuit depth and optimization method, we set $U$ equal to the exact unitary that diagonalizes the Hamiltonian, while still varying over the spin effective energies $\{ \epsilon_r \}$. Using this construction to calculate TFD correlation functions, we display our results for $N=20$ at $T=0.2$ in Figure \ref{fig:SYK2TFDCorrelation}.

We see that Figure \ref{fig:SYK2TFDCorrelation} indicates good agreement between the exact and approximate TFD correlation functions for a variety of $\alpha$ over this time range, with very little deviations between these quantities. We attribute the accuracy of this scenario to the use of the exact unitary that diagonalizes the Hamiltonian, as this matrix necessarily captures the behavior of the exact solution. The PSA is therefore capable of producing accurate results, although the optimal unitary transformation may be rather complex for certain Hamiltonians. The failure of the simplest optimization method to find a good approximation may be due to insufficient circuit depth or to an inadequacy in our gate-by-gate optimization strategy.

\subsubsection{Comparison to an alternative method}
In Reference \cite{maldacena2018eternal}, an alternative method is proposed to prepare approximate thermofield double states for the SYK model. This method postulates a Hamiltonian that acts on both the left ($L$) and right ($R$) copies of the system and whose ground state approximates the thermofield double state. Explicitly, the Hamiltonian is
\begin{equation}
\begin{split}
H_{\text{total}} = H_{\text{L,SYK}}+H_{\text{R,SYK}} + H_{\text{int}}, \\
H_{\text{int}} = i \mu \sum_{j=1}^N \chi_j^L \chi_j^R. \quad \quad \quad
\end{split}
\end{equation}
Here, $H_{\text{L,SYK}}$ and $H_{\text{R,SYK}}$ are SYK Hamiltonians on the left and right subsystems. $H_{\text{int}}$ is an interaction Hamiltonian that couples Majorana fermions on the left subsytem ($\chi^L_j$) with those on the right subsystem ($\chi^R_j$), parameterized by $\mu \in \mathbb{R}$. For small values of $\mu$, we expect the ground state of this system, $|G\rangle$, to approximate the TFD. More details on this construction can be found in References \cite{maldacena2018eternal,gu2017spread}.

\begin{figure}[h]
\begin{subfigure}{.24\textwidth}
  \centering
  \includegraphics[width=4.5cm]{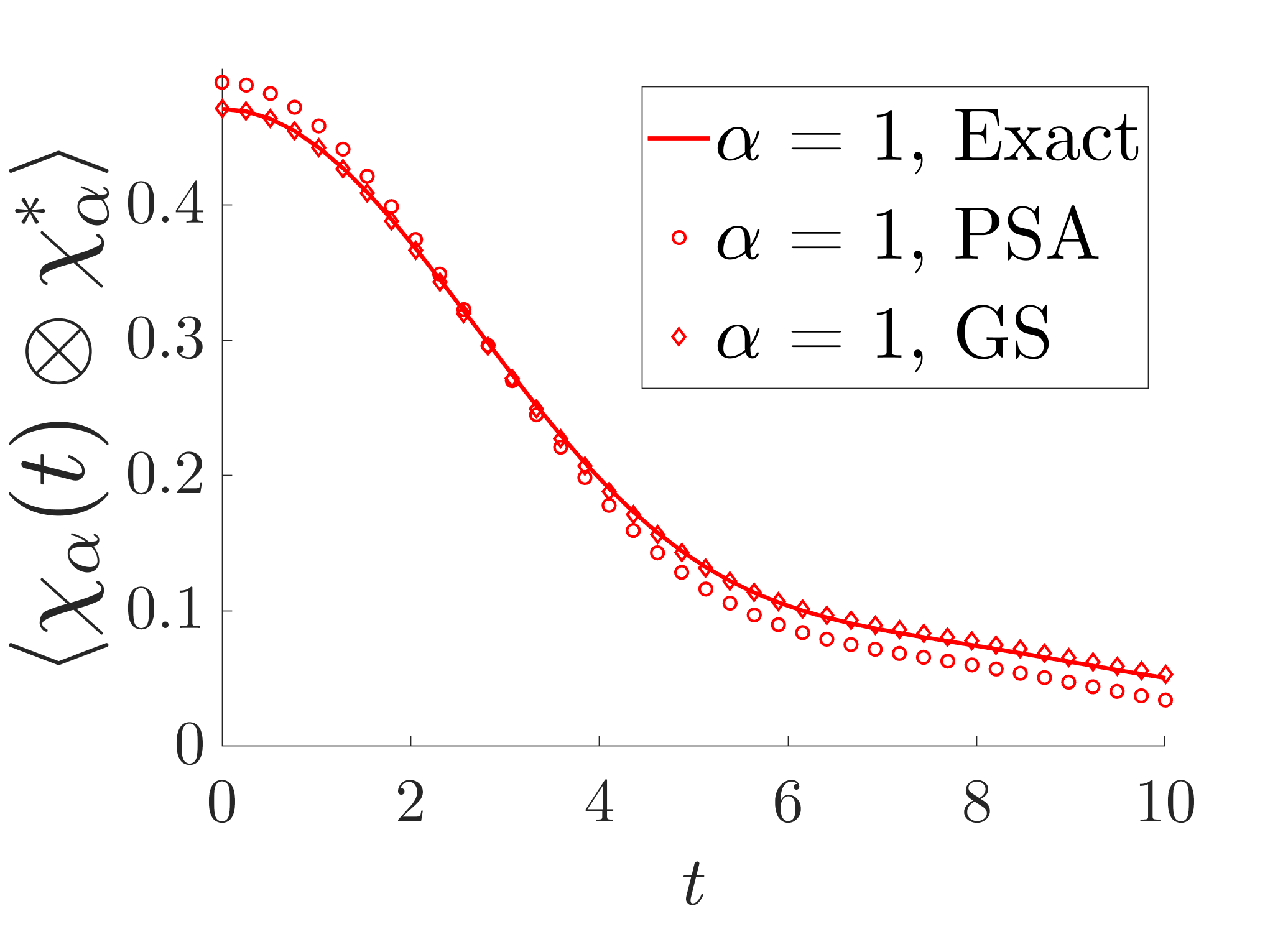}
  \label{fig:SYK_TFD_Comparison_1}
\end{subfigure}%
\begin{subfigure}{.24\textwidth}
  \centering
  \includegraphics[width=4.5cm]{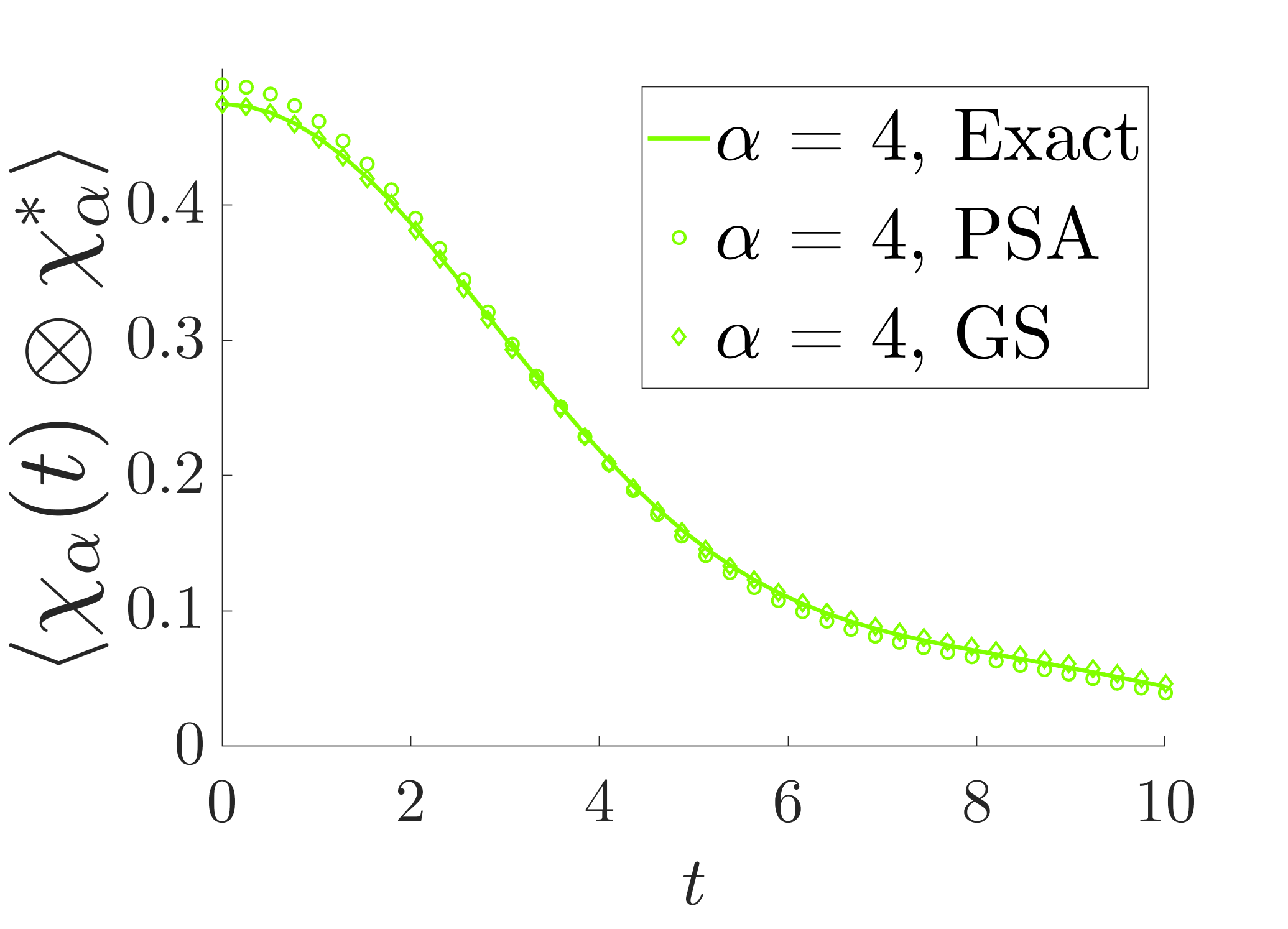}
  \label{fig:SYK_TFD_Comparison_4}
\end{subfigure}
\begin{subfigure}{.24\textwidth}
  \centering
  \includegraphics[width=4.5cm]{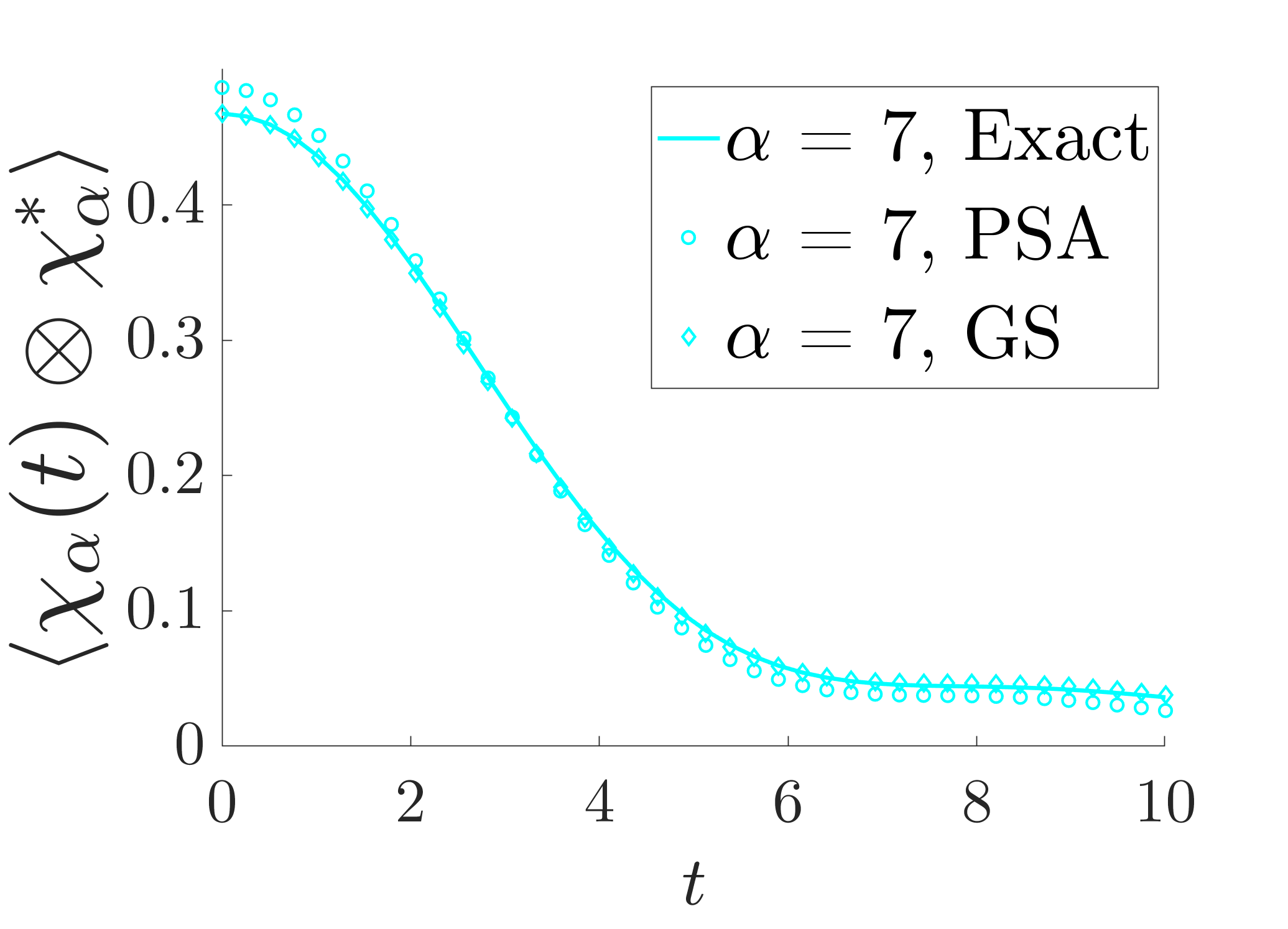}
  \label{fig:SYK_TFD_Comparison_7}
\end{subfigure}%
\begin{subfigure}{.24\textwidth}
  \centering
  \includegraphics[width=4.5cm]{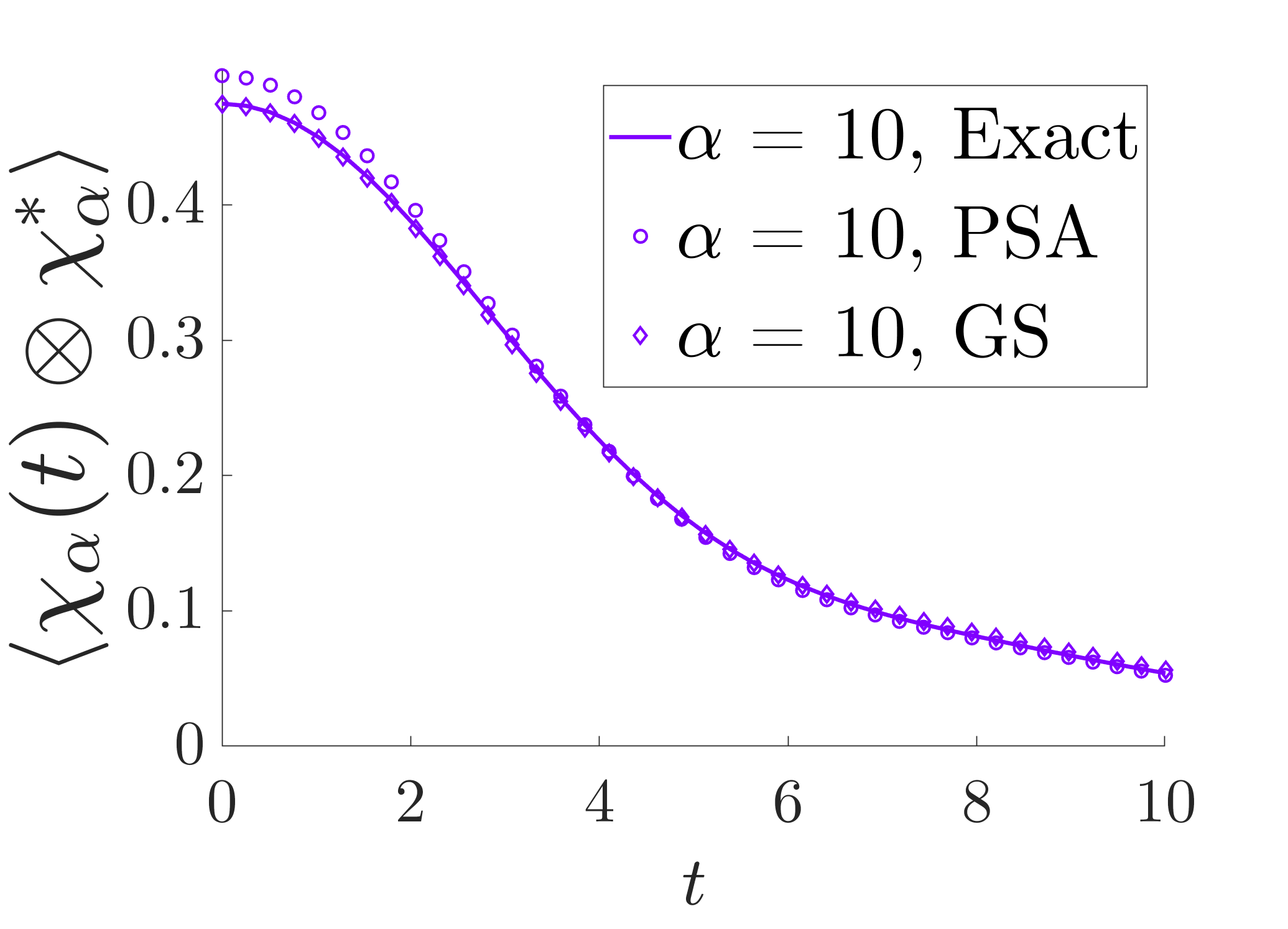}
  \label{fig:SYK_TFD_Comparison_10}
\end{subfigure}%
\caption{Exact, PSA, and ground state method (GS) TFD correlation functions $\langle \chi_\alpha (t) \otimes \chi_\alpha^* \rangle$ for the SYK model with $N=12$, $T=0.5$, and $\ell=10$. $\alpha \in\{ 1,4,7,10\}$ are displayed as representatives of all correlation functions.}
\label{fig:SYKTFDCorrelationComparison}
\end{figure}

We now compare this alternative ground state method to the PSA. We implement this procedure by determining the optimum value of $\mu$ such that $|G \rangle$ well approximates the TFD. Following Reference \cite{maldacena2018eternal}, we choose the value of $\mu$ that enforces $\langle G | H_{\text{L,SYK}}| G \rangle = \langle TFD | H_{\text{L,SYK}}| TFD \rangle$. Using this procedure, we consider the scenario with $N=12$ Majorana fermions in the $L$ and $R$ subsystems each at $T=0.5$. To compare the two methods, we calculate the TFD correlation functions $\langle \chi_\alpha (t) \otimes \chi_\alpha^* \rangle$ with the PSA and with the alternative methods, which we will call the ground state method. Our result are plotted against the corresponding exact correlations functions in Figure \ref{fig:SYKTFDCorrelationComparison}.

The plots in Figure \ref{fig:SYKTFDCorrelationComparison} clearly demonstrate that the correlation functions of the PSA TFD and the ground state of $H_{\text{total}}$ agree well with the exact TFD correlation functions, being nearly indistinguishable for most $\alpha$. However, Figure \ref{fig:SYKTFDCorrelationComparison} also indicates the ground state method performs better than the PSA in some scenarios. In particular, we note the the PSA correlation functions show slight deviations from the exact correlation functions at early times ($t\lesssim 2$), whereas the correlation functions of the ground state agree well even at early times. In addition, in the $\alpha = 1$ and $\alpha = 7$ cases, we see that the PSA correlation functions deviate from the exact values more than the the ground state correlation functions at late times ($t \gtrsim 5$). Nevertheless, as the PSA and ground state correlation functions still follow the same trends as the exact correlation functions, both approaches capture the dynamics of the TFD.

\section{Discussion}
We have illustrated that the PSA applies well to the production of approximate thermal states and TFD states for both spin chains and the SYK model. The properties of the PSA thermal states are seen to agree with the exact solution across various models, couplings, and temperatures. From the properties of the PSA discussed in Section \ref{sec:Relations}, our results demonstrate that the precise details of an energy spectrum are not needed to reproduce ordinary thermal observables. Hence, one need not know the exact spectrum of a Hamiltonian to construct its approximate thermal state. In addition, as our construction revolves around the application of unitaries to a product state, it follows that a variety of quantum thermal states can be efficiently approximated on quantum computers, wherein one applies nearest neighbor unitary gates to a product state of qubits. In Appendix \ref{sec:appendixA}, we provide further justification for the PSA. Starting from the eigenstate thermalization hypothesis, discussed in Reference \cite{ETH}, we argue that thermal observables calculated with the TFD state of the PSA should agree well with the exact TFD observables provided the exact Hamiltonian-diagonalizing unitary is used.

As emphasized above, our approach does not require knowledge of the full many-body energy spectrum and is variational. A key feature which underlies the utility of the variational approach is that the entropy can be efficiently computed and the energy can be efficiently measured (or computed). Another virtue of the PSA is that it is systematically improvable. The obvious refinement parameter is the circuit depth, since with a large enough depth any unitary can be well approximated. To refine the spectrum, instead of considering product states of single spins, one could instead consider product states of clusters of spins. As long as the clusters are not too large, the entropy will be efficiently computable and thermofield double states efficiently preparable. More generally, one has a family of approximations in which Markov states are used to generate the spectrum. Such states have the property that their entropies can be efficiently computed from local data, and their corresponding thermofield double states can be efficiently prepared. We expect this generalization is sufficient for local Hamiltonians since a wide class of thermal states of interest are known to be approximate quantum Markov states.

Another key question is whether the PSA can be implemented on near term quantum devices. The present discussion assumed the ability to apply arbitrary quantum gates with perfect fidelity, which will only be possible on a fully fault-tolerant quantum computer. However, it may be that arbitrary unitary transformations are not required for a specific Hamiltonian, so that if one had access to a restricted but natural set of unitaries (for example, the ability to evolve in time with the Hamiltonian of interest and the ability to apply local fields), it might still be possible to well approximate the thermal state. Investigating this question in more detail in different near term quantum computers and quantum simulators is an interesting direction for future work. In addition, studying the properties of the PSA under realistic noise conditions is also quite interesting. We have not yet been able to prove the kind of noise resilience expected for ground state tensor networks~\cite{2017arXiv171107500K}, so this is also an instructive direction for followup work.

For local spin chains, we found that the PSA worked exceedingly well and that the minimization of the free energy could be carried out straightforwardly. For the SYK model, we found that the PSA worked well down to very low temperatures provided the exact diagonalizing unitary was used. At high temperatures, the PSA with a low depth circuit well approximated correlations in the thermal state and the TFD state, but at lower temperatures, with the same depth circuit, the PSA struggled to precisely reproduce TFD correlations, although correlations in the mixed thermal state were reasonably well captured. Hence, the exact unitary calculation provides a proof of principle that the PSA can capture SYK correlations, but more work is needed at low temperatures to find a relatively low depth PSA approximation. One expects the requisite circuit depth to increase with decreasing temperature, so it may simply be that a larger depth is required. However, another possibility is that the free energy minimization is more difficult at low temperatures in the SYK model. We currently minimize the free energy gate-by-gate, but a more global approach, where multiple gates are varied simultaneously, might give a better result given the all-to-all nature of the interactions in this model. It may also be useful to minimize the free energy at multiple temperatures simultaneously using a fixed variational energy spectrum. For example, the appearance of $\rho^{1/2}$, which involves a significant enhancement of the low probability states in $\rho$, in TFD correlations suggests the need to accurately capture correlations at $T$ and $T/2$. In any event, we expect that the full complexity of the exact diagonalizing unitary is not needed to reproduce thermal correlations, so with some additional work on the optimization side, it seems likely that the low temperature SYK limit can also be well-captured with a modest depth circuit.

The ability to prepare approximate thermal states in experiments will be valuable in a wide variety of contexts. One particularly interesting direction is to carry out experiments on toy models of quantum gravity, such as the SYK model or AdS/CFT duality. In those contexts, the TFD state is dual to a particularly simple geometry describing two entangled black holes, a geometry which underlies many simple probes of quantum gravity, ranging from scrambling to teleportation through wormholes~\cite{maldacena2018eternal, Almheiri:2012rt, Almheiri:2013hfa,susskind2016er, Gao:2016bin, maldacena2017diving}. Hence, one can envision first verifying that classical gravity well describes the dynamics of interest, then using a quantum simulator or quantum computer to study the system in novel regimes beyond our current understanding. States that arise from measuring half of a thermofield double states in a particular basis have even been suggested to give insights into the interiors of black holes~\cite{Kourkoulou:2017zaj,Almheiri:2018ijj,deBoer:2018ibj} and potentially provide toy models of quantum cosmology~\cite{Cooper:2018cmb}. To that end, another important direction arising from our work is to explore the degree to which the PSA TFD state reproduces the physics of the exact TFD state for the above purposes.

\textit{Acknowledgements}: We thank Shenglong Xu for helpful discussions and for suggesting to us Reference \cite{PhysRevB.79.144108}. We also thank Mohammad Hafezi and Alireza Seif for early discussions on related work and Sagar Lokhande for feedback on the manuscript. This work is supported by the Simons Foundation via the It From Qubit Collaboration.

\textit{Note}: After this work was completed but a few days before it appeared, two other independent works studying a similar problem appeared~\cite{Cottrell:2018ash,Wu:2018nrn}.

\bibliography{APS_Formatted_Refs}
\bibliographystyle{apsrev4-1}

\appendix

\section{Product spectrum ansatz and the eigenstate thermalization hypothesis} \label{sec:appendixA}

The eigenstate thermalization hypothesis (ETH) states that for observables chosen from a privileged class, matrix elements between energy eigenstates ``look like" random variables of a special type. Given an observable $O$, the mathematical statement of ETH is
\be \label{eq:eth}
\langle E_a | O |E_b \rangle = f_O(\bar{E})\delta_{ab} + e^{-S(\bar{E})/2} g_O(\bar{E},\omega) R^O_{ab},
\ee
where $\bar{E}=\frac{E_a+E_b}{2}$ and $\omega= E_a-E_b$ and $R^O_{ab}$ is a random variable with mean zero and unit variance. The functions $f_O$ and $g_O$ are smooth functions of their arguments, and $f_O$ is just the microcanonical average of $O$. This hypothesis can be used to justify the product spectrum ansatz. The size of the system, $N$, is proportional to the number of particles or volume or some extensive thermodynamic variable. The physical meaning of the randomness assumption is that when computing physical observables requiring a sum over many energy eigenstates, one can reliably evaluate the leading behavior of such sums by replacing the summands with random variables as outlined above.

Given an operator $V$ in one copy of the system, let $V_L = V \otimes I$ and $V_R= I \otimes V$ denote the action of $V$ on the left and right, respectively, of the thermofield double system. The transpose in the energy basis is denoted $V^{\mathbb{T}}$. Consider the correlation function
\be
G = \langle \text{TFD} | V_L V^{\mathbb{T}}_{R} | \text{TFD} \rangle - \langle V \rangle \langle V \rangle,
\ee
for some Hermitian $V$. In terms of the energy basis it is
\bea
G = & \sum_{a,b} \sqrt{p_a p_b} \langle E_a | V| E_b\rangle \langle E_b | V | E_a \rangle \nonumber \\
& - \sum_{a,b} p_a p_b \langle E_a | V |E_a \rangle \langle E_b | V | E_b \rangle,
\eea
where $p_a = \frac{e^{-\beta E_a}}{Z}$ is the Boltzmann weight. Using the ETH ansatz in Eq.~\eqref{eq:eth}, the correlator $G$ consists of three terms proportional to $f g$, $g f$, and $g^2$. The $f^2$ term approximately cancels assuming that $f$ and $g$ are smoothly varying functions of the energy density $\bar{E}/N$, since the distribution over energy densities induced by $p_a$ is sharply peaked with fluctuations of order $1/\sqrt{N}$.

The goal is to compare the exact value of $G$ to an approximate value obtained by replacing $p_a$ with a product spectrum distribution. Let the product spectrum distribution be $q_a$, and suppose it has the property that the corresponding energy distribution has the same peak value and variance as the true distribution $p_a$.

Using the PSA thermofield double state, we define an approximate $G_{\text{PSA}}$. By the same argument as above, the $f^2$ terms still cancel since we assumed $q_a$ induces a sharply peaked distribution of energy density. Suppose for simplicity that $\langle V \rangle =0$, so that $f$ is zero. Then the only term to consider is
\be
\sum_{ab} \sqrt{p_a p_b} g(\bar{E},\omega)^2 e^{-S(\bar{E})}R_{ab} R_{ba}
\ee
in the exact case, and
\be
\sum_{ab} \sqrt{q_a q_b} g(\bar{E},\omega)^2 e^{-S(\bar{E})} R_{ab} R_{ba}
\ee
in the approximate case. Hermiticity of $V$ gives $R_{ba}=R_{ab}^*$, so both terms contain only $|R_{ab}|^2$.

Let $\phi_{ab} = g^2 e^{-S} |R_{ab}|^2$; this quantity is clearly positive and of order $1/\mathcal{D}$ where $\mathcal{D}=e^{S}$ is the effective number of states. In addition, let $\Delta_{ab} = \sqrt{p_a p_b} - \sqrt{q_a q_b}$; we expect $\Delta_{ab}$ to oscillate rapidly as a function of $ab$ and to be effectively uncorrelated with $R_{ab}$. As $q_a$ and $p_a$ produce similar well peaked energy distributions and have most of their support on the same set of states, $q_a$ cannot systematically deviate from $p_a$. Hence, because $p_a$ and $q_a$ are of order $1/\mathcal{D}$, we expect $\Delta_{ab} \sim \pm 1/\mathcal{D}$. The difference between $G$ and $G_{\text{PSA}}$ is then
\be
\delta G := \sum_{ab} \Delta_{ab} \phi_{ab}.
\ee
By the arguments given above, $\delta G$ should be of order $1/\mathcal{D}$ due to random walk scaling. As such, we expect $\delta G \rightarrow 0$ in the thermodynamic limit. This justifies the use of the product spectrum ansatz for calculating thermal observables of many-body systems.

\end{document}